\documentclass[a4paper, 12pt, oneside]{article}

\usepackage{aapreprint}

\usepackage{graphicx,wrapfig,float,slashed,subcaption,bbold,bm}
\usepackage{amsmath,amssymb,epsfig,graphicx,xcolor}
\usepackage{epstopdf}
\usepackage{ragged2e}

\newcommand{\nc}{\newcommand}
\nc{\non}{\nonumber}
\nc{\hc}{\hbox {H.c.}}
\nc{\noi}{\noindent}
\nc{\barx}{\bar{x}}
\nc{\pbarn}{\;\hbox {pb}}
\nc{\fbarn}{\;\hbox {fb}}

\nc{\hsp}{\hspace{0.5cm}}
\nc{\lsp}{\hspace{1cm}}
\nc{\Lsp}{\hspace{2cm}}
\nc{\LLsp}{\lsp\lsp}
\nc{\lra}{\longrightarrow}
\nc{\p}{\prime}
\nc{\sgn}{\text{sgn}}
\nc{\ph}{\varphi}
\nc{\op}{{\cal O}}

\nc{\beq}{\begin{equation}}  \nc{\eeq}{\end{equation}}
\nc{\bea}{\begin{eqnarray}}  \nc{\eea}{\end{eqnarray}}
\nc{\baa}{\begin{array}}     \nc{\eaa}{\end{array}}
\nc{\bit}{\begin{itemize}}   \nc{\eit}{\end{itemize}}
\nc{\ben}{\begin{enumerate}} \nc{\een}{\end{enumerate}}
\nc{\bce}{\begin{center}}    \nc{\ece}{\end{center}}
\nc{\bpm}{\begin{pmatrix}}   \nc{\epm}{\end{pmatrix}}
\nc{\bvt}{\begin{verbatim}}  \nc{\evt}{\end{verbatim}}

\def\lsim{\mathrel{\raise.3ex\hbox{$<$\kern-.75em\lower1ex\hbox{$\sim$}}}}
\def\gsim{\mathrel{\raise.3ex\hbox{$>$\kern-.75em\lower1ex\hbox{$\sim$}}}}

\def\udots{\mathinner{\mkern1mu\raise1pt\vbox{\kern7pt\hbox{.}}\mkern2mu\raise4pt\hbox{.}\mkern2mu\raise7pt\hbox{.}\mkern1mu}}

\def\mev{\;\hbox{MeV}}

\allowdisplaybreaks

\newcommand\fverb{\setbox\fverbbox=\hbox\bgroup\verb}
\newcommand\fverbdo{\egroup\medskip\noindent%
			\fbox{\unhbox\fverbbox}\ }
\newcommand\fverbit{\egroup\item[\fbox{\unhbox\fverbbox}]}
\newbox\fverbbox

\preprint{\begin{flushright}
MITP/18-113\\
UTTG-25-18
\end{flushright}}

\title{Collider Signals of the Mirror Twin Higgs through 
the Hypercharge Portal} 
 \author[a]{Zackaria Chacko,} 
\author[b]{Can Kilic,}
\author[c,d]{Saereh Najjari,}
\author[e]{and Christopher B. Verhaaren}
\affiliation[a]{Maryland Center for Fundamental Physics, Department of Physics,\\ University of Maryland, College Park, MD 20742-4111 USA}
\affiliation[b]{Theory Group, Department of Physics\\ University of Texas at Austin,
Austin, TX 78712, USA}
\affiliation[c]{Theoretische Natuurkunde \& IIHE/ELEM,\\ Vrije Universiteit Brussel, Pleinlaan 2, 1050 Brussels, Belgium}
\affiliation[d]{PRISMA Cluster of Excellence \& Mainz Institute for Theoretical Physics,\\ Johannes Gutenberg University, 55099 Mainz, Germany}
\affiliation[e]{Center for Quantum Mathematics and Physics (QMAP), Department of Physics,\\ University of California, Davis, CA, 95616-5270 USA.}
\emailAdd{zchacko@physics.umd.edu}
\emailAdd{kilic@physics.utexas.edu}
\emailAdd{saereh.najjari@vub.be}
\emailAdd{cbverhaaren@ucdavis.edu}

\abstract{
 We consider the collider signals arising from kinetic mixing between 
the hypercharge gauge boson of the Standard Model and its twin 
counterpart in the Mirror Twin Higgs model, in the framework in which 
the twin photon is massive. Through the mixing, the Standard Model 
fermions acquire charges under the mirror photon and the mirror $Z$ 
boson. We determine the current experimental bounds on this scenario, 
and show that the mixing can be large enough to discover both the twin 
photon and the twin $Z$ at the LHC, or at a future 100 TeV hadron 
collider, with dilepton resonances being a particularly conspicuous 
signal. We show that, in simple models, measuring the masses of both the 
mirror photon and mirror $Z$, along with the corresponding event rates 
in the dilepton channel, overdetermines the system, and can be used to 
test these theories.
 }


\begin{document}

\maketitle
\flushbottom

\section{Introduction}
\label{Introduction}

 With the discovery of the Higgs boson at the Large Hadron Collider 
(LHC)~\cite{Aad:2012tfa,Chatrchyan:2012xdj}, all the particles predicted 
by the Standard Model (SM) have now been observed, and the search for 
new physics is well underway. Among the many puzzles this program may 
shed light on is the large hierarchy between the mass of the Higgs and 
the Planck scale. If a symmetry protects the Higgs mass from short 
distance quantum effects, then we expect new particles with masses of 
order the weak scale that are related to the SM particles by the 
symmetry. The symmetry partners of the top quark, the top partners, are 
expected to be particularly light. Despite increasingly sophisticated 
efforts, however, such particles have yet to be discovered at the LHC.

The paradigm of neutral 
naturalness~\cite{Chacko:2005pe,Barbieri:2005ri,Chacko:2005vw,Burdman:2006tz,Poland:2008ev,Cai:2008au,Craig:2014aea,Craig:2014roa,Batell:2015aha,Csaki:2017jby,Serra:2017poj,Cohen:2018mgv,Cheng:2018gvu,Dillon:2018wye,Xu:2018ofw} 
addresses the hierarchy problem by incorporating color neutral top 
partners. Since color neutral particles are much more difficult for the 
LHC to discover, the bounds on this class of models remain relatively 
weak. In particular, the Mirror Twin Higgs (MTH) 
framework~\cite{Chacko:2005pe} protects the Higgs mass by employing 
fermionic top partners that are singlets under all the SM gauge 
groups.\footnote{Recently, SM-singlet \emph{scalar} top partner models 
have also been constructed~\cite{Cohen:2018mgv,Cheng:2018gvu} .} This 
protection results from a global symmetry in the Higgs sector combined 
with a discrete $\mathbb{Z}_2$ symmetry that exchanges each SM field 
with a mirror (``twin") copy that is charged under its own gauge groups. 
The electroweak (EW) gauge symmetries of both the SM and twin sectors 
are contained in the approximate global symmetry, which in the simplest 
realization of the model is $SU(4)\times U(1)$. This global symmetry is 
spontaneously broken at a scale $f$ down to $SU(3)\times U(1)$. The 
longitudinal modes of the $W$ and $Z$ vector bosons and the physical 
Higgs boson are among the resulting pseudo-Nambu-Goldsone bosons 
(pNGBs). The mass of the Higgs is protected against large radiative 
corrections by a combination of the non-linearly realized global 
symmetry and the discrete $\mathbb{Z}_2$ twin symmetry.

After electroweak symmetry breaking the Higgs and its twin counterpart 
mix. If the $\mathbb{Z}_2$ symmetry were exact, the scales of 
electroweak symmetry breaking in the SM sector and the twin sector would 
be identical, $v= v'$, where $v = 246$ GeV is the Higgs vacuum 
expectation value (VEV) in the SM, and we employ primes to denote the twin 
sector. Then, as a consequence of the mixing, the couplings of the Higgs 
boson to SM states would be suppressed and it would have equal couplings 
to both visible and twin states. This, however, conflicts with existing 
experimental results on the couplings of the Higgs. Introducing soft 
breaking of the discrete $\mathbb{Z}_2$ symmetry allows the Higgs VEV 
$v$ to be small compared to $v'$, so that the couplings of the physical 
Higgs boson to visible sector particles are close to their SM values, 
while its couplings to twin particles are suppressed. In this framework 
the ratio $v/f \equiv v/\sqrt{v^2 + {v'}^2}$ determines many 
observables, such as the ratio of the masses of the SM particles and 
their twin partners, as well as deviations in the couplings of the Higgs 
boson away from their SM values. The soft-$\mathbb{Z}_2$ breaking also 
introduces a tuning in the model, which scales as $(v/f)^2$. 
Experimental bounds on the Higgs couplings constrain $f/v\gtrsim 3$, 
while requiring the model be less than 10\% tuned indicates $f/v\lesssim 
6$~\cite{Burdman:2014zta}. This provides a definite window for the twin 
particle masses: between 3 and 6 times the mass of their SM 
counterparts.

The MTH framework has been explored and expanded on in recent years. It 
has been shown that the breaking of the discrete $\mathbb{Z}_2$ symmetry 
can be realized 
spontaneously~\cite{Beauchesne:2015lva,Yu:2016bku,Yu:2016swa,Jung:2019fsp,Batell:2019ptb}. 
Ultraviolet completions have been constructed based on 
supersymmetry~\cite{Falkowski:2006qq,Chang:2006ra,Craig:2013fga,Katz:2016wtw, 
Badziak:2017syq,Badziak:2017kjk,Badziak:2017wxn} (see 
also~\cite{Berezhiani:2005ek}), compositeness of the 
Higgs~\cite{Geller:2014kta,Barbieri:2015lqa,Low:2015nqa}, and most 
recently using a ``turtle" construction~\cite{Asadi:2018abu}. Composite 
MTH models have been shown to be consistent with precision electroweak 
constraints~\cite{Contino:2017moj} and flavor 
bounds~\cite{Csaki:2015gfd}, and their collider signals 
studied~\cite{Cheng:2015buv,Cheng:2016uqk}.

The cosmology of the MTH, in its original incarnation, is problematic. 
In the early universe, Higgs-mediated interactions keep the mirror 
sector in thermal equilibrium with the SM down to temperatures of order 
a few GeV~\cite{Barbieri:2005ri}. Then the twin photon and twin 
neutrinos give an overly large contribution to the total energy density 
in radiation, leading to conflict with the bounds on dark radiation from 
the cosmic microwave background and Big Bang nucleosynthesis. This 
problem can be solved if the model is extended to incorporate an 
asymmetric reheating process that contributes to the energy density in 
the SM degrees of freedom, but not to the mirror 
sector~\cite{Berezhiani:1995yi,Berezhiani:1995am}. To solve the problem, 
this mechanism must operate at late times, after the two sectors have 
decoupled. This can be realized without requiring additional breaking of 
the discrete $Z_2$ symmetry that relates the two 
sectors~\cite{Chacko:2016hvu,Craig:2016lyx}. An alternative approach is 
to introduce hard breaking of the $Z_2$ into the twin sector Yukawa 
couplings, thereby altering the spectrum of mirror 
states~\cite{Farina:2015uea,Barbieri:2016zxn,Csaki:2017spo,Barbieri:2017opf}. 
This affects the number of degrees of freedom in the two sectors at the 
time of decoupling, allowing the cosmological bounds to be satisfied. 
Once this problem has been resolved, questions such as the nature of 
dark matter~\cite{Farina:2015uea,Cheng:2018vaj}, the origin of the 
baryon asymmetry~\cite{Farina:2016ndq,Earl:2019wjw}, the order of the 
electroweak phase transition~\cite{Fujikura:2018duw}, and the 
implications of the MTH framework for large scale 
structure~\cite{Chacko:2018vss}, can be addressed.

The cosmological challenges can also be solved by making the twin sector 
vector-like~\cite{Craig:2016kue}. An even more radical approach is to 
simply remove from the theory the two lighter generations of mirror 
fermions and the twin photon, which do not play a role in solving the 
little hierarchy problem. This construction, known as the Fraternal Twin 
Higgs (FTH)~\cite{Craig:2015pha}, gives rise to distinctive signatures at the LHC involving 
displaced vertices~\cite{Curtin:2015fna,Csaki:2015fba}, and admits 
several promising dark matter 
candidates~\cite{Craig:2015xla,Garcia:2015loa,Garcia:2015toa,Hochberg:2018vdo,Terning:2019hgj}.

In the MTH framework the discrete $\mathbb{Z}_2$ symmetry and the 
resulting gauge invariance under [SU(2) $\times$ U(1)]$^2$ largely 
sequester the SM fields from their twin counterparts. The Higgs boson 
itself is the only low-energy portal between the two sectors that is 
guaranteed by construction. However, if the UV completion is weakly 
coupled, the radial mode associated with the breaking of the symmetry 
can be light enough to probe at current and future 
colliders~\cite{Ahmed:2017psb,Chacko:2017xpd,Buttazzo:2018qqp,Kilic:2018sew,Alipour-fard:2018mre}. 
In the absence of additional new fields~\cite{Bishara:2018sgl}, there is 
only one other renormalizable operator consistent with the symmetries 
that can link the two sectors; a kinetic mixing of the SM hypercharge 
gauge boson $B_\mu$ with its twin counterpart $B_\mu'$,
 \begin{equation}
\frac{\epsilon}{2}B_{\mu\nu}'B^{\mu\nu}.
\label{mixepsilon}
 \end{equation}
 Here $B_{\mu\nu}$ is the usual Abelian field strength. This operator 
has the effect of giving the SM quarks and leptons charges proportional 
to $\epsilon$ under the twin photon and twin $Z$, which are denoted $A'$ 
and $Z'$ respectively.

Apart from some discussion in the context of dark 
matter~\cite{Craig:2015xla,Garcia:2015toa}, previous analyses have 
largely neglected this mixing, since it is not radiatively generated in 
the low-energy theory to at least three-loop 
order~\cite{Chacko:2005pe,Low:2015nqa}. An $\epsilon$ generated at the 
four-loop level is small enough to remain consistent with the very 
strong bounds on the photon mixing with another massless 
vector~\cite{Davidson:2000hf,Vogel:2013raa}; for a twin electron of mass 
of order MeV these bounds require $\epsilon\lesssim 10^{-9}$. However, 
ultraviolet completions of the MTH based on compositeness generically 
introduce states charged under both SM and twin gauge groups, with the 
result that $\epsilon$ is expected to receive one-loop corrections of 
order $10^{-2}$\textendash$10^{-3}$. Such large values of $\epsilon$ can 
only be accommodated if the twin photon $A'$ is massive.

 In this paper we consider the collider signals arising from kinetic 
mixing between the hypercharge gauge boson of the SM and its twin 
counterpart in the MTH model, in the framework in which the twin photon 
is massive. We focus on the framework in which the discrete 
$\mathbb{Z}_2$ symmetry is only softly broken. We determine the current 
experimental bounds on this scenario, and explore the possibility of 
discovering both the twin photon and the twin $Z$ at the LHC, or at a 
future 100 TeV hadron collider such as the Future Circular Collider in 
hadron mode (FCC-hh). Although the phenomenology of new vector bosons 
that mix with the U(1$)_{\rm Y}$ of the SM has been thoroughly 
explored~\cite{Holdom:1985ag,Babu:1997st,ArkaniHamed:2008qp,Abel:2008ai,Goodsell:2009xc,Goodsell:2011wn,Essig:2013lka,Hoenig:2014dsa,Curtin:2014cca}, 
the MTH framework introduces several novel features.

A massive twin photon requires contributions to the masses of the mirror 
gauge bosons beyond those from electroweak symmetry breaking. This can 
be accomplished either by simply introducing a mass term for the twin 
hypercharge gauge boson, or by extending the Higgs sector. In our 
analysis, we consider both cases. For the first case, we simply include 
an explicit Proca mass term for the twin hypercharge boson,
 \begin{equation}
\frac{m_{B'}^2}{2}B_\mu'B'^\mu,
\label{eq:Bpmass}
 \end{equation}
 where $m_{B'}$ is a free parameter in the Lagrangian. Since a Proca 
mass term can be obtained from the St\"{u}ckelberg construction, which 
is unitary and renormalizable, after gauge fixing, this theory is 
ultraviolet complete. The Proca mass can also be thought of as arising 
from the VEV of a new Higgs field that carries charge under twin 
hypercharge, but not under twin SU(2)$_{\rm L}$~\cite{Batell:2019ptb}. If this Higgs field is 
heavy, it will decouple from the low energy spectrum, but if it carries 
only a small charge under U(1$)_{\rm Y'}$, the twin photon, though no 
longer massless, will still be light.

Alternatively, rather than include an explicit mass term for $B_{\mu}'$, 
the twin photon can acquire a mass from the VEV of an additional Higgs 
field in the twin sector. For concreteness, we consider a scenario in 
which the Higgs content of the theory is extended to include two Higgs 
doublets in each of the SM and twin 
sectors~\cite{Chacko:2005vw,Beauchesne:2015lva,Yu:2016bku,Yu:2016swa}. 
The VEVs of the two doublets in the visible sector must be aligned to 
leave the photon massless, but the twin sector doublets are not so 
constrained. In particular, the VEVs of the two Higgs doublets in the 
twin sector must be somewhat misaligned to ensure that the twin photon 
has a mass.

As a consequence of the coupling in Eq.~(\ref{mixepsilon}), the massive 
$Z'$ and $A'$ mix with the hypercharge gauge boson. Therefore the SM 
quarks and leptons acquire charges under the twin photon and twin $Z$, 
which can therefore be searched for at colliders. At hadron colliders, 
production through light quarks followed by decay into dileptons is a 
particularly promising channel. We find that values of the mixing 
parameter $\epsilon$ as small as $10^{-2}$ can potentially be probed at 
the LHC and FCC-hh.

In the softly-broken $\mathbb{Z}_2$ scenario, a few physical parameters 
determine all the couplings in the model that are relevant for LHC 
searches. In particular, in the models we consider, the collider signals 
depend on only three parameters beyond those in the SM. Consequently, if 
more than three independent measurements can be performed, the 
parameters of the model are overdetermined, allowing these
theories to be tested. Measuring the masses of the $A'$ and $Z'$, along with 
both their dilepton resonance production rates, achieves this goal.

In the following section
 we outline the interactions of the neutral vector bosons in the MTH 
framework, and study their production rates and branching fractions. 
This is done for two scenarios; a model in which the twin hypercharge 
gauge boson is given an explicit Proca mass (THPM), and a twin two Higgs 
doublet model (T2HDM). In Sec.~\ref{sec:constraints} we determine the 
existing bounds on these two models from direct searches and from 
precision measurements. Then, in Sec.~\ref{s.discovery}, we explore the 
discovery reach of the high luminosity LHC (HL-LHC) and a future 100 TeV 
hadron collider, and consider the prospects for testing the THPM and the 
T2HDM as the underlying origin of these resonances. We conclude in 
Sec.~\ref{s.Con}.

\section{Neutral Vectors in the Mirror Twin Higgs}
\label{Mirror Twin Higgs Models}

In this section we study the effects of kinetic mixing on the masses and 
couplings of the neutral vector bosons in the MTH model, in the scenario 
in which the twin photon is massive. We then explore the implications of 
mixing for the production cross sections and decay widths of these 
particles at colliders. We focus on the two case studies of a Proca mass 
for the twin hypercharge gauge boson and a two Higgs doublet version of 
the MTH model. For each case we obtain the production cross sections and 
branching fractions of the electrically neutral twin gauge bosons $A'$ 
and $Z'$ at the LHC and at a future 100 TeV collider. 

\subsection{Proca Mass for Twin Hypercharge}

In this first case, we begin by considering the ${\rm SU(2)}_{\rm L} \times 
{\rm U}(1)_{\rm Y} \times {\rm SU(2)}_{\rm L'}\times {\rm U(1)}_{\rm Y'}$ 
gauge sector,
 \begin{align}
\mathcal{L}\supset&-\frac{1}{2}{\rm Tr}\left[{W_{\mu\nu}}{W}^{\mu\nu}\right]-\frac{1}{2}{\rm Tr}\left[{W_{\mu\nu}^{\p}}{W^{\p}}^{\mu\nu}\right]-\frac{1}{4}{B_{\mu\nu}}{B}^{\mu\nu}-\frac{1}{4}{B_{\mu\nu}^{\p}}{B^{\p}}^{\mu\nu}\nonumber\\
&+\frac{\epsilon}{2 }{B_{\mu\nu}^{\p}}B^{\mu\nu}+\frac{m^2_{B'}}{2}B^{\p}_{\mu}B^{\p\mu},
\label{Lagrangian}
 \end{align}
 where primed fields belong to the twin sector and $\epsilon$ is the 
kinetic mixing parameter. For any spin-1 field $X_{\mu}$, we employ the 
notation
 \beq
X_{\mu\nu}=\partial_\mu X_\nu-\partial_\nu X_\mu,
 \eeq
 to denote the field strength, with the usual generalization to the
non-Abelian case. 

Rather than work directly with $B'$ and $W'^{3}$, it is convenient to go 
over to the analogue of the familiar electroweak basis of the photon and 
$Z$ in the twin sector. Writing $s_W \equiv \sin\theta_W$ and $c_W 
\equiv \cos\theta_W$, where $\theta_W$ is the weak mixing angle, we 
define
 \begin{equation}
\overline{Z}_{\mu}^{\p} \equiv c_WW_\mu^{\p 3}-s_WB_{\mu}^\p,
\ \ \ \
\overline{A}_{\mu}^\p \equiv s_WW_\mu^{\p 3}+c_WB_{\mu}^\p \;.
 \end{equation}
 We find that, as expected, the Proca mass induces mass mixing between 
the two vector bosons,
 \beq
\frac12 \left(\overline{A}_{\mu}^\p\;\; \overline{Z}_{\mu}^{\p} \right)\left( \begin{array}{cc}
c_W^2m^2_{B'} & -s_W c_W m^2_{B'}\\
-s_W c_W m^2_{B'} &m^2_{\overline{Z}^\p}+ s_W^2m^2_{B'}
\end{array}\right)\left(\begin{array}{c}
\overline{A}_{\mu}^\p\\
\overline{Z}_{\mu}^{\p}
\end{array} \right).\label{e.PMTHMassMat}
 \eeq
 Here $m^2_{\overline{Z}^\p} \equiv (v'/v)^2 m^2_{Z_0}$, where 
$m_{Z_0}$ is the mass of the $Z$ boson in the SM. This mass matrix for 
the neutral twin sector vector bosons can be diagonalized by a simple 
rotation of the fields, with the mixing angle given by
 \begin{align}
 \sin2\phi=\frac{m^2_{B'}s_{2W}}{\sqrt{\left( m^2_{B'}-m^2_{\overline{Z}^\p}\right)^2+4s_W^2m^2_{B'}m^2_{\overline{Z}^\p}}}\,.
\label{mixangle1}
 \end{align}
 We label the lighter of the two mass eigenstates as $A_{\mu}'$, and the 
heavier one as $Z_{\mu}'$. Note that in the limit $m^2_{B'} \gg 
m^2_{\overline{Z}^\p}$, Eq.~(\ref{mixangle1}) implies that 
$\phi=\theta_W$. Essentially, this undoes the weak mixing, meaning that 
for large $m_{B'}$, the mass eigenstates are, to a good approximation, 
simply the $W_3'$ and the $B'$ with masses $m_{\overline{Z}^\p} c_{W}$ 
and $m_{B'}$ respectively. In this limit the $Z'$, which has its mass 
set by $m_{B'}$, couples more strongly to the visible sector than the 
lighter $A'$, which decouples from the SM as $m_{B'}$ increases. 
Additional details are given in Appendix~\ref{a.Diag}.
 
In the opposite limit, $m^2_{B'} \ll m^2_{\overline{Z}^\p}$, the mixing angle 
$\phi$ tends to zero. The mass eigenstates are then, to a good 
approximation, simply $\overline{A}_{\mu}^{\p}$ and $\overline{Z}_{\mu}^{\p}$, with masses 
$m_{B'} c_W$ and $m_{\overline{Z}^\p}$, respectively. In this limit it is the 
lighter eigenstate, the $A'$, which couples more strongly to the visible 
sector, while the couplings of the heavier $Z'$ are suppressed by a 
relative factor of $\tan\theta_W$.

\begin{figure}
\centering
\includegraphics[trim={8mm 11mm 8mm 8mm},clip,width=0.6\textwidth]{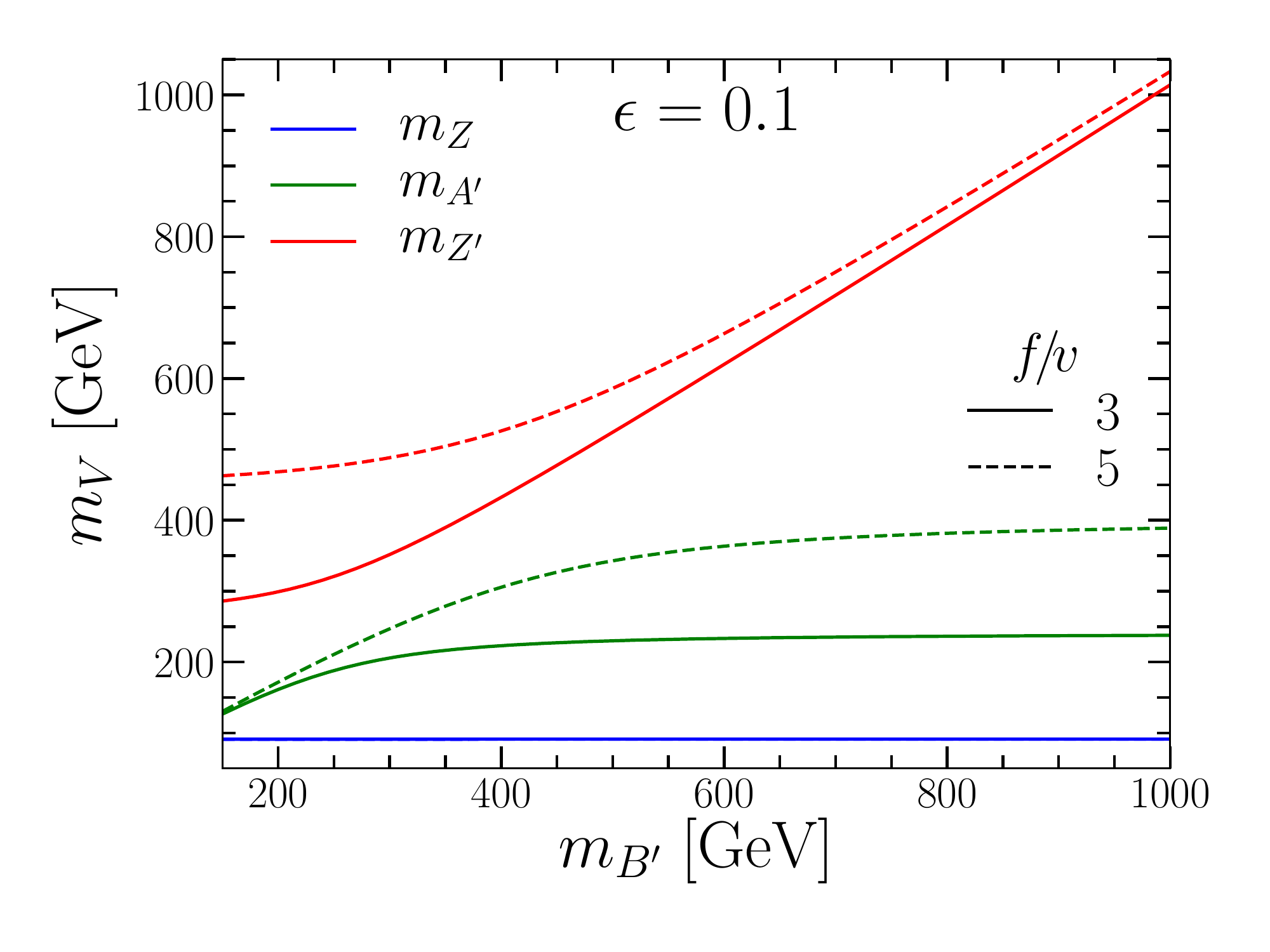}
\caption{The mass eigenvalues for the twin neutral vector bosons as a function of the Proca mass $m_{B'}$ in the THPM model for $\epsilon=0.1$ and $f/v=3$ and 5.}
\label{fig:masseigenvalues}
\end{figure}

To analyze the experimental signals, we must transform from 
Eq.~\eqref{Lagrangian} to a basis diagonal in both mass and kinetic 
terms. This is accomplished by first performing a shift of $B$ and then 
rotating into the mass basis. The details of this procedure are given in 
Appendix~\ref{a.Diag}, where we obtain the field transformation to 
leading order in $\epsilon$. The couplings of the diagonalized fields to 
SM and twin particles are provided in Appendix~\ref{a.VFcouplings}, 
again to leading order in $\epsilon$. However, this perturbative 
analysis breaks down near mass degeneracies, where one of the twin 
sector gauge bosons becomes close in mass to the SM $Z$ boson. Therefore 
the results we present have been obtained by diagonalizing the system 
numerically. For small $\epsilon$, away from mass degeneracies, the mass 
eigenvalues are close to the values obtained by diagonalizing the matrix 
Eq.~(\ref{e.PMTHMassMat}). The mass eigenvalues as a function of 
$m_{B'}$, for the benchmark values $f/v=3$ and 5 and $\epsilon=0.1$, are 
shown in Fig.~\ref{fig:masseigenvalues}. We see that 
for large $m_{B'}$, the mass of the heavier $Z'$ is close to $m_{B'}$, 
while that of the lighter $A'$ asymptotes to $(v'/v) m_W$.
\begin{figure}
\centering
\includegraphics[trim={8mm 11mm 8mm 8mm},clip,width=0.49\textwidth]{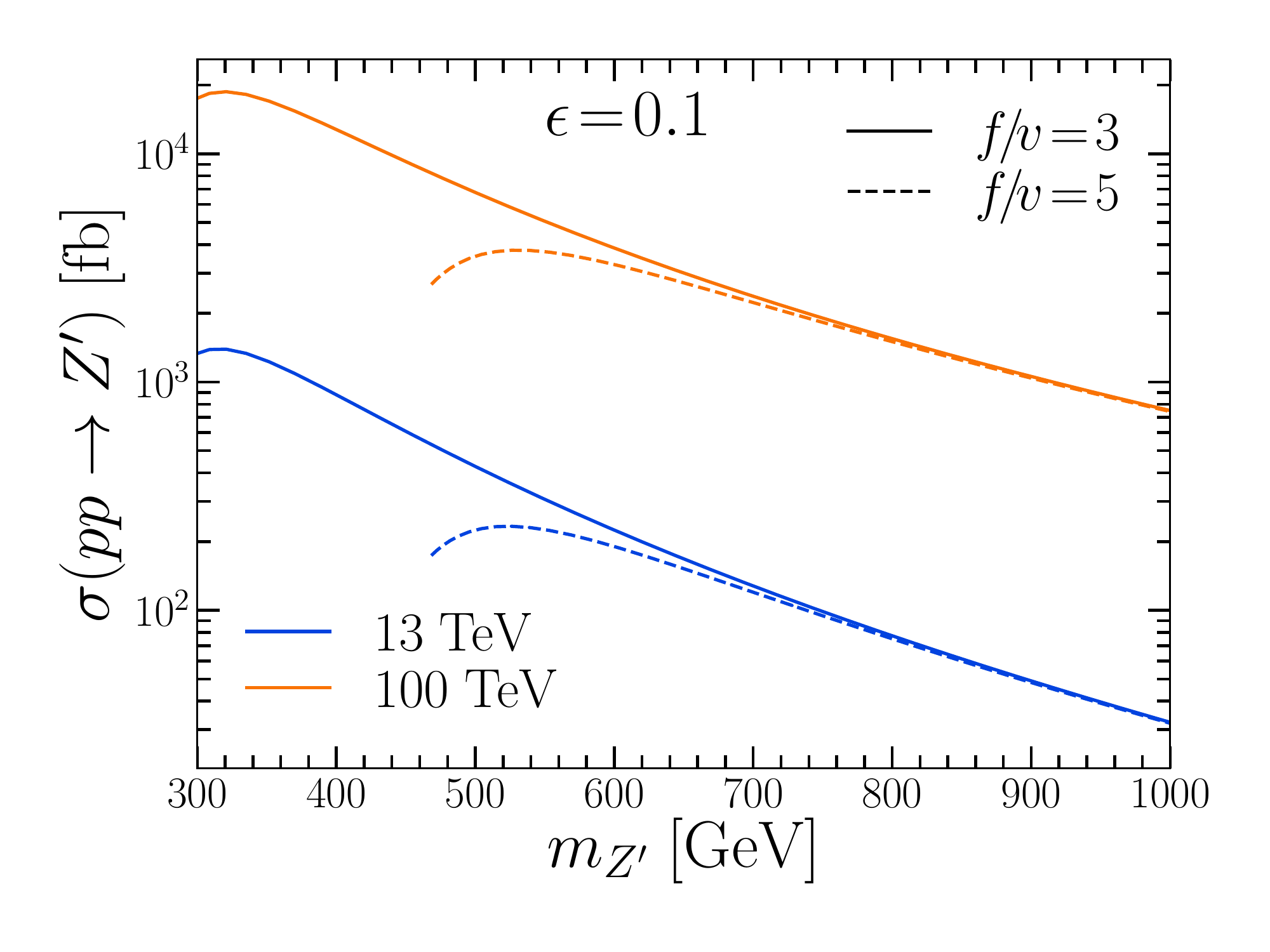}
\includegraphics[trim={8mm 11mm 8mm 8mm},clip,width=0.49\textwidth]{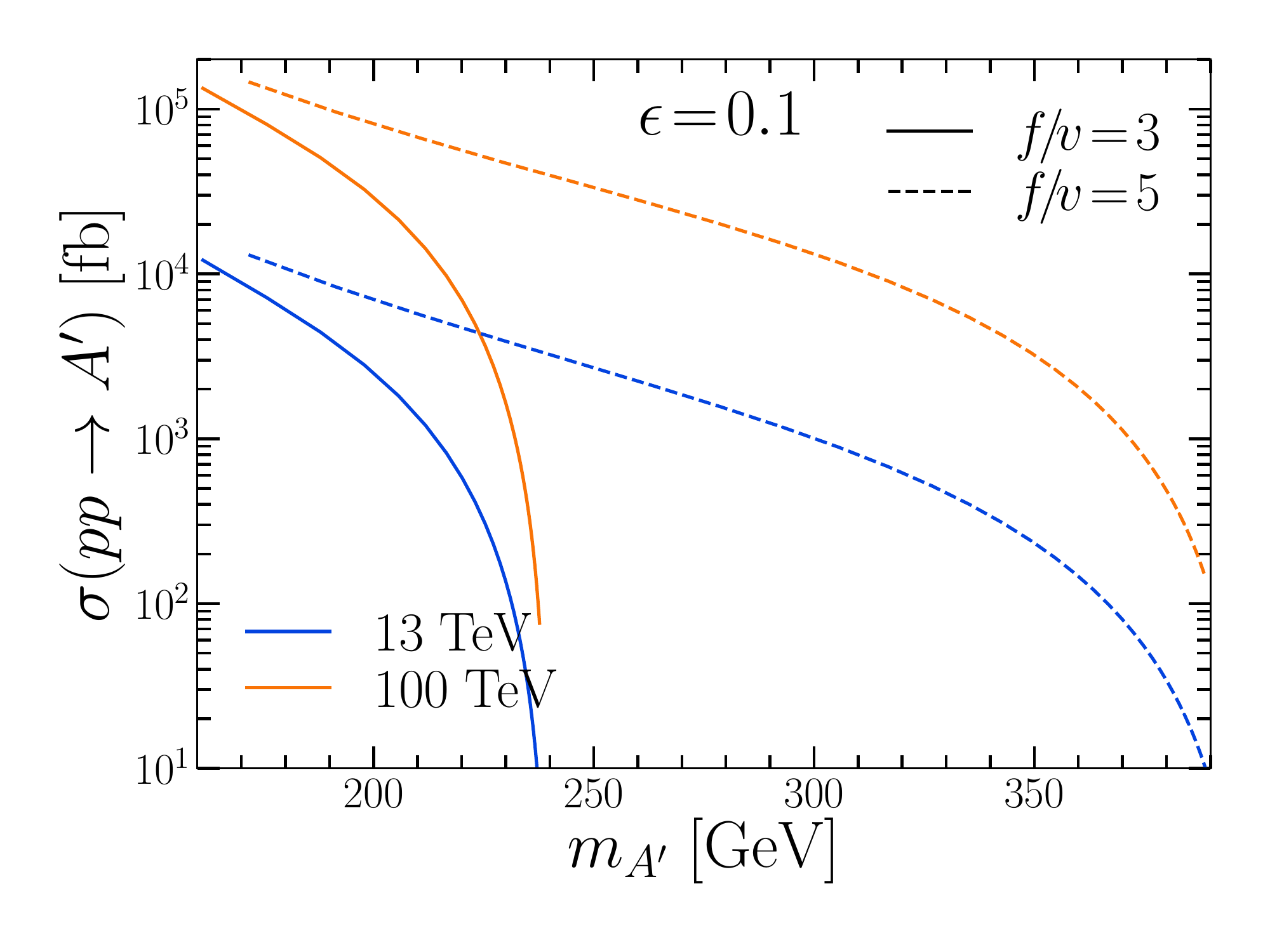}
\caption{Production cross section of the $Z'$ (left) and $A'$ (right) in the THPM model at the LHC and at a future 100~TeV hadron collider for $\epsilon=0.1$ and $f/v=3$ and 5.}
\label{fig:productionxsec}
\end{figure}

After diagonalization the SM fermions acquire charges under the $A'$ and 
the $Z'$. This allows the twin vector bosons to be produced and detected 
at colliders. The production cross sections of the $A'$ and $Z'$ at the 
13 TeV LHC and at a 100~TeV future hadron collider are shown in 
Fig.~\ref{fig:productionxsec}, for $\epsilon=0.1$ and $f/v = 3$ and 5. 
We see that for small values of its mass the $A'$ cross section is 
large, but drops off quickly as the mass increases. This occurs because, 
for large $m_{B'}$, the $A'$ is almost entirely composed of $W_3'$, 
which does not mix directly with SM hypercharge. Consequently, as 
$m_{B'}$ increases, the cross section plummets. In contrast, the 
production cross section of the $Z'$ remains sizable even for masses in 
the TeV range.
\begin{figure}
\centering
\includegraphics[trim={8mm 10mm 8mm 8mm},clip,width=0.49\textwidth]{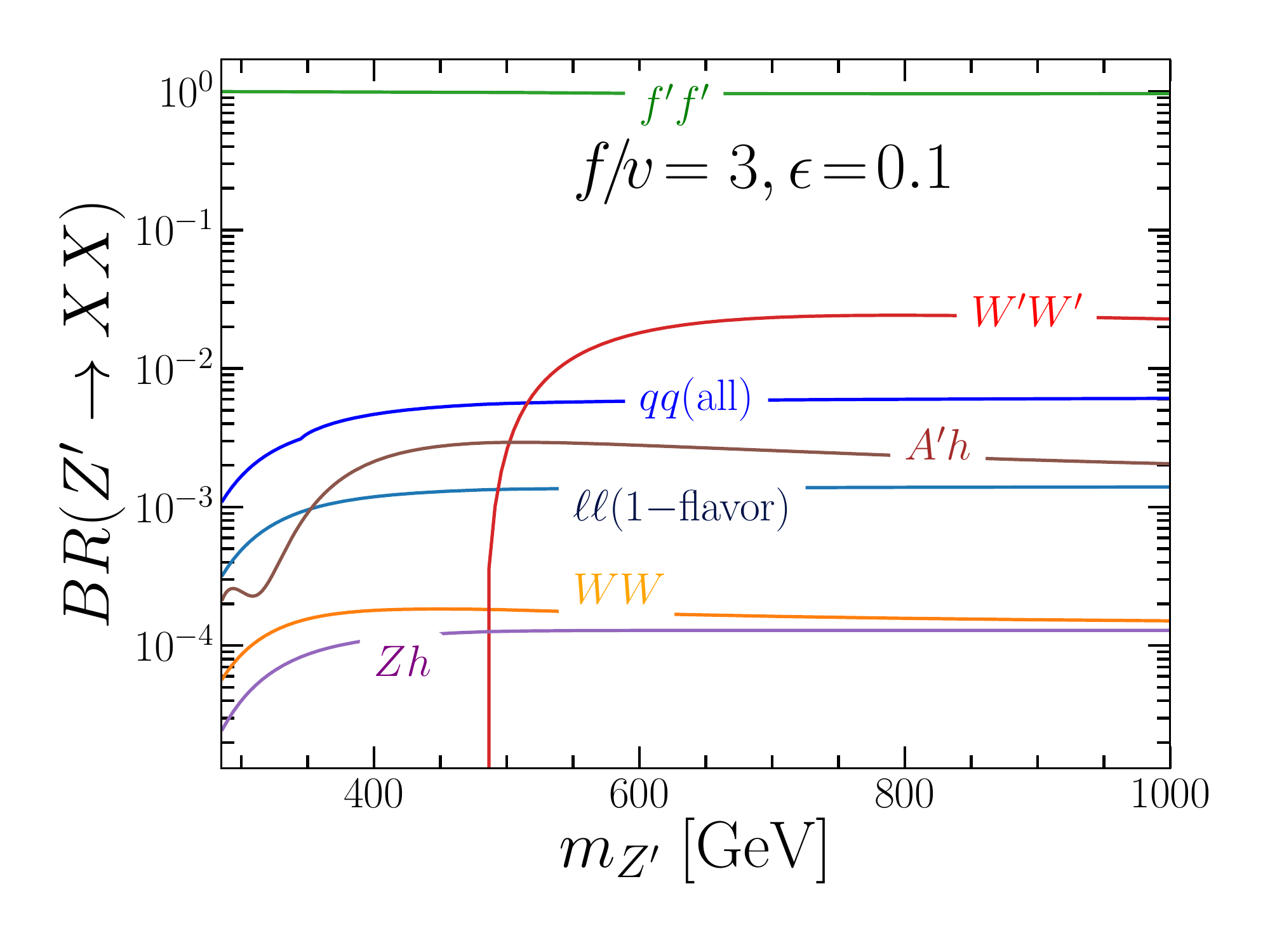}
\includegraphics[trim={8mm 10mm 8mm 8mm},clip,width=0.49\textwidth]{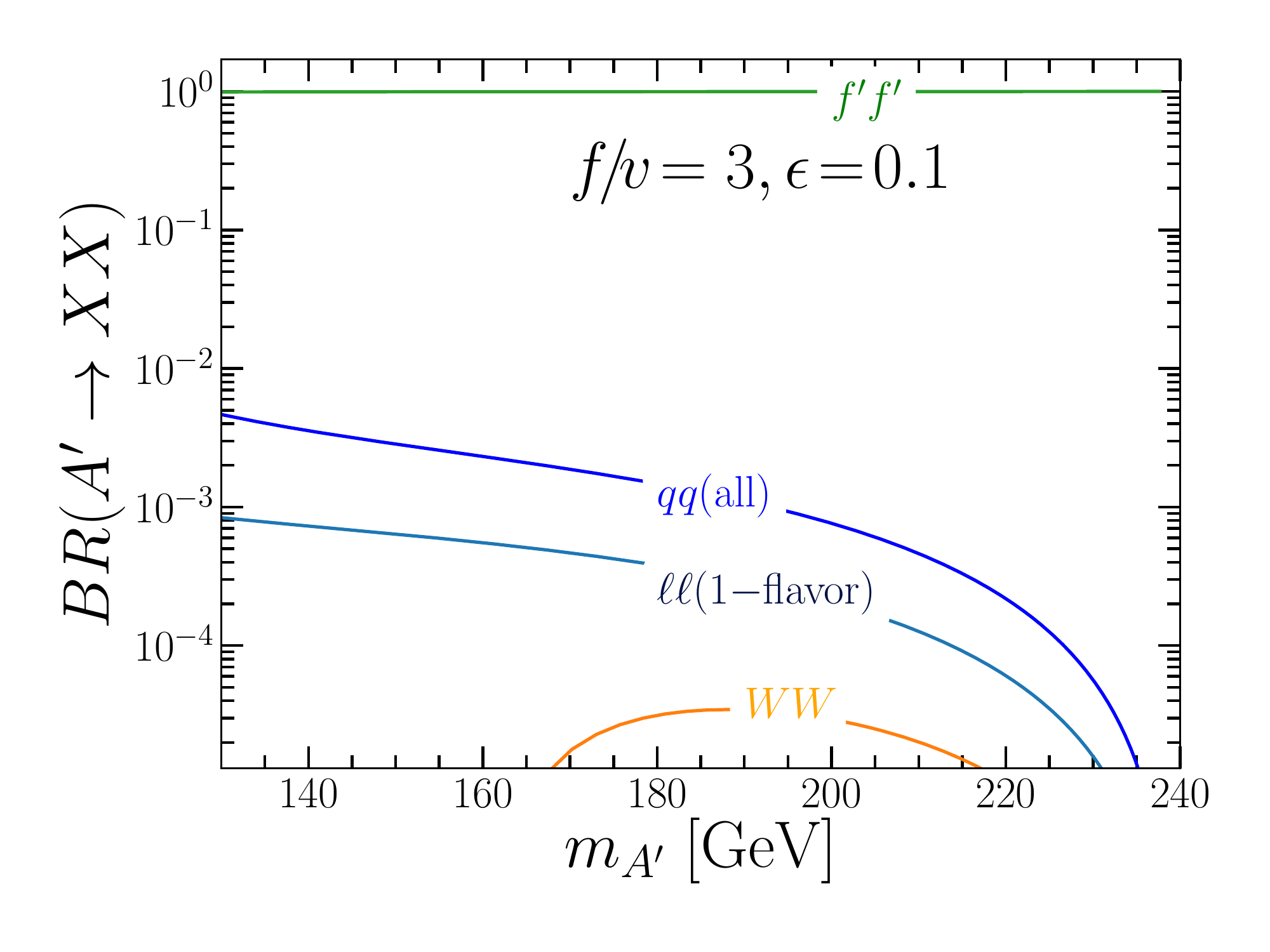}\\
\includegraphics[trim={8mm 11mm 8mm 8mm},clip,width=0.49\textwidth]{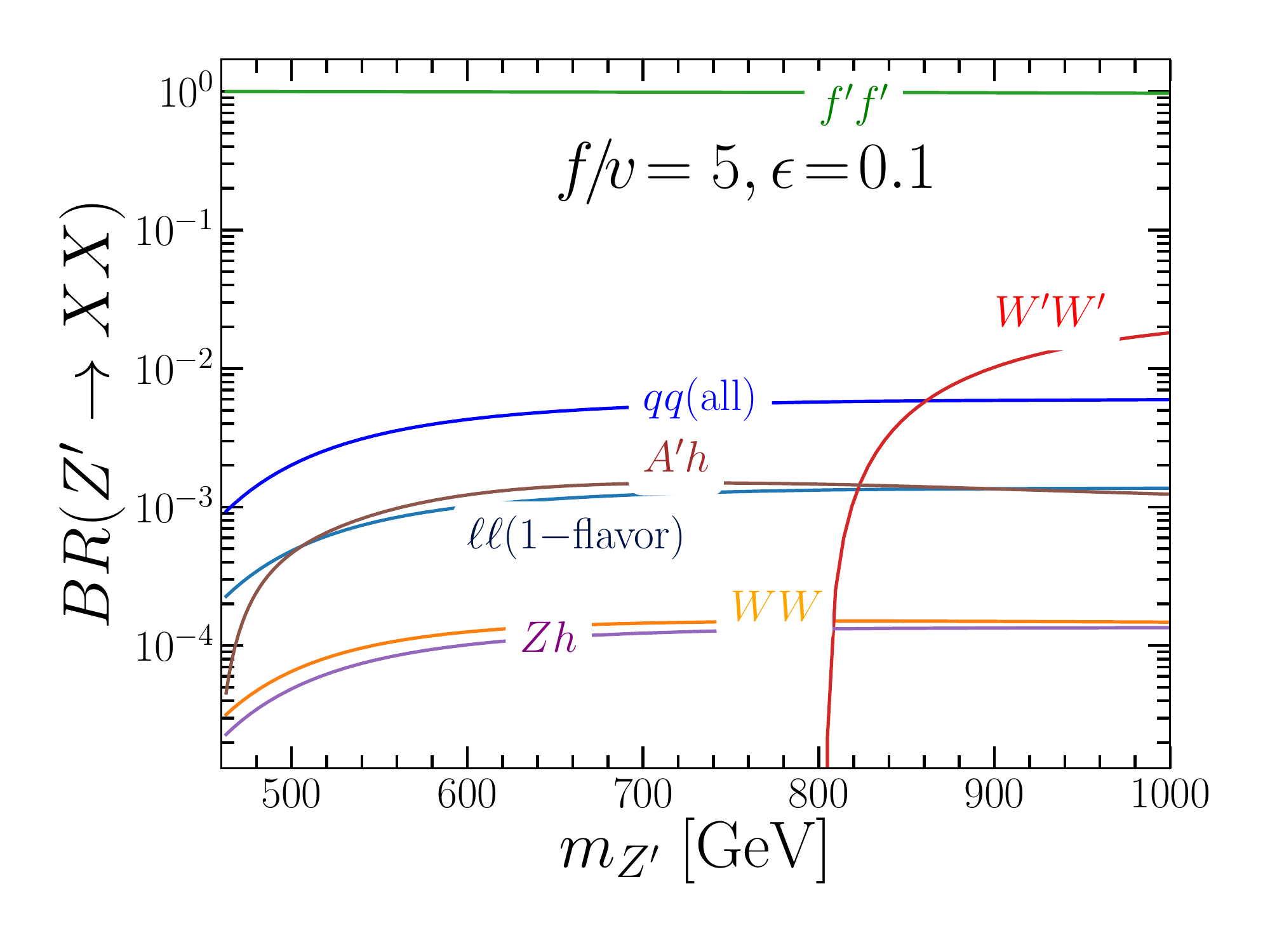}
\includegraphics[trim={8mm 11mm 8mm 8mm},clip,width=0.49\textwidth]{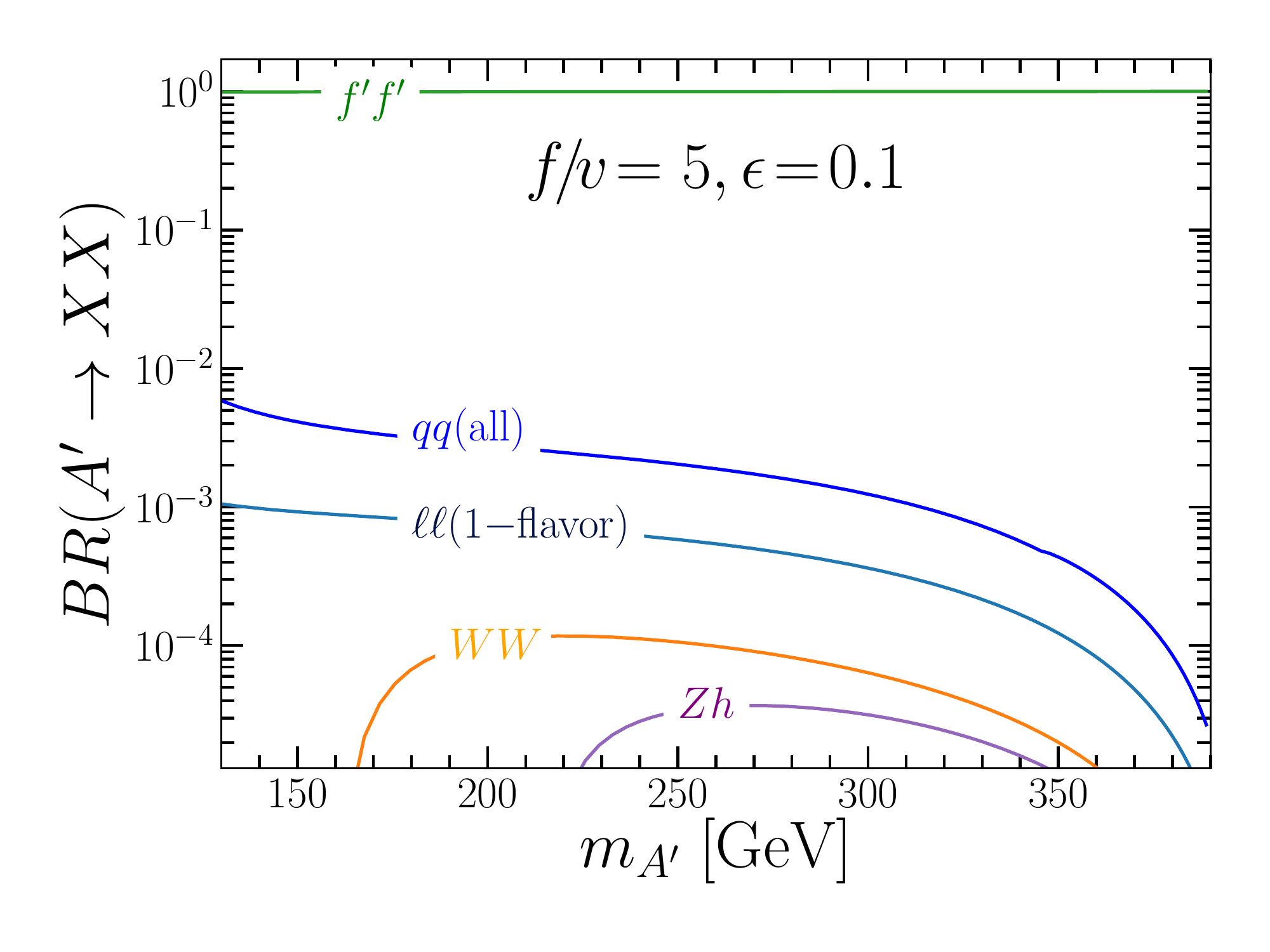}
\caption{Branching ratios of the $Z'$ (left) and $A'$ (right) for the THPM model for $\epsilon=0.1$ and $f/v=3$(5) for the top (bottom) row. The light blue curve is for a single SM lepton flavor, while the dark blue curve corresponds to all six quark flavors.}
\label{fig:BRs}
\end{figure}

The corresponding branching fractions are shown in Fig.~\ref{fig:BRs}. 
As expected, decays to twin fermions dominate the width. Nevertheless, 
the branching fractions into SM fermions can be large enough for 
discovery in a clean resonant channel such as dileptons. Note that for 
large values of $m_{B'}$, the branching fraction of $A'$ into dileptons 
is much smaller than the corresponding $Z'$ branching fraction. These 
hierarchies in the production cross sections and dilepton branching 
ratios translate into a much greater discovery potential for the $Z'$ at 
colliders in most of parameter space, as we show below.

\subsection{Twin Two Higgs Doublet Model\label{T4HDM}}

We now turn to the twin two Higgs doublet model. We label the two 
visible sector Higgs doublets as $H_1$ and $H_2$, and their twin 
counterparts as $H_1'$ and $H_2'$. Since the photon is massless, the 
VEVs of $H_1$ and $H_2$ must be aligned. However, the VEVs of the 
doublets in the twin sector need to be misaligned if they are to give 
mass to both the neutral gauge bosons in the twin sector.

The alignment of the two twin doublet VEVs can be quantified in many ways. 
For instance, the quantity
 \begin{equation}
\frac{|H^{'\dag}_{1}H^{'}_{2}|^{2}}  {(H^{'\dag}_{1}H^{'}_{1})(H^{'\dag}_{2}H^{'}_{2})},
 \end{equation}
 tends to one in the limit of perfect alignment, and to zero in the 
limit of perfect anti-alignment. The limit of perfect alignment does not 
serve our purposes since it leaves one vector boson massless, but any 
other configuration will result in both the neutral gauge bosons in the 
twin sector acquiring masses. For concreteness, we restrict our 
attention to the limit of perfect anti-alignment in the twin sector. We 
further assume that of the two visible sector doublets, only $H_1$ 
acquires a VEV, while $\langle H_2 \rangle = 0$. It is straightforward 
to construct a potential for the scalar sector that gives rise to these 
features. While the potential for $H_2$ and $H_2'$ must respect the 
discrete $\mathbb{Z}_2$ twin symmetry, it need not obey the global SU(4) 
$\times$ U(1) symmetry. Therefore the scalar states in $H_2$, the second 
visible sector doublet, are not required to be pNGBs and can 
therefore naturally be heavy.

The fact that both $H_1'$ and $H_2'$ acquire VEVs implies that there are 
two additional pNGB states in the mirror sector. If these 
states are to acquire a mass, there must be interactions that couple 
$H_1'$ and $H_2'$ in the potential for the scalar fields. We can lift 
these additional pNGBs from the low energy theory by 
including the term
 \begin{equation}
V= 	\lambda \left(\left| {H_1^\dagger  H_2}\right|^2+\left| {H'^\dagger_1  H'_2}\right|^2\right),\label{e.scalarMasses}
 \end{equation}
 in the scalar potential. These terms respect the discrete 
$\mathbb{Z}_2$ symmetry, and therefore do not generate a large mass for 
the light Higgs. For $\lambda \gtrsim 1$, these states acquire masses 
above those of the twin gauge bosons, and we can neglect their dynamics 
at low energies.

We can then write the VEVs of the Higgs fields as,  
 \begin{equation}
H_1=\left( \begin{array}{c}
0\\
v/\sqrt{2}
\end{array}
\right), \ \ \ \  
H'_1=\left( \begin{array}{c}
0\\
 f_1\cos\vartheta/\sqrt{2}
\end{array}
\right),\ \ \ \
H_2=\left( \begin{array}{c}
0\\
0
\end{array}
\right),
 \ \ \ \
H_2'=\left( \begin{array}{c}
\displaystyle f_2/\sqrt{2} \\
0
\end{array}
\right),
\label{e.linearTwinFourHiggsDoubletsAsector}
 \end{equation}
where $\sin\vartheta=v/f_1$. The mass terms for the gauge bosons arise from the kinetic terms for the Higgs bosons,
 \begin{align}
{\cal L}& \supset\left| {D}_{\mu} {H_1}\right|^2+\left| {D}_{\mu} H_2\right|^2+\left|D'_{\mu} H'_1\right|^2+\left| D'_{\mu} H'_2\right|^2~.\label{Lagrangiantwin}
 \end{align}
 From this we obtain the usual mass terms for the visible sector gauge bosons,
 \beq
\frac{g^2v^2}{4}W^+_\mu W^{-\mu}+\frac{g^2v^2}{8c_W^2}Z_{\mu} Z^\mu~.\label{e.AvecBosons}
 \eeq
 The mass matrix for the twin sector is more complicated. In this case we 
find
 \begin{align}
&\frac{g^2}{4}\left(f_1^2\cos^2\vartheta+f_2^2 \right)W'^+_\mu W'^{-\mu}+\frac{g^2}{2}s_W^2f_2^2\overline{A}'_{\mu} \overline{A}'^\mu+\frac{g^2s_W}{2c_W}f_2^2c_{2W}\overline{A}'_{\mu} \overline{Z}'^\mu\nonumber\\
&+\frac{g^2}{8c_W^2}\left[f_1^2\cos^2\vartheta+f_2^2c_{2W}^2 \right]\overline{Z}'_{\mu} \overline{Z}'^\mu\, .
\end{align}
 where $\cos n\theta_W\equiv c_{nW}$. It is convenient to define 
$m_{\overline{Z}^\p}=m_{Z_0}\cot\vartheta$ where $m_{Z_0}$ is the $Z$ 
mass in the SM and $m_{\overline{A}'}=gf_2s_W$. Expressed in terms of 
these variables, the mass matrix is given by ($t_{2W}\equiv\tan 2\theta_W$)
 \beq
\frac12 \left(\overline{A}_{\mu}^\p\;\; \overline{Z}_{\mu}^{\p} \right)\left( \begin{array}{cc}
m_{\overline{A}'}^2 & m_{\overline{A}'}^2/t_{2W}\\
m_{\overline{A}'}^2/t_{2W} &m^2_{\overline{Z}^\p}+ m_{\overline{A}'}^2/t_{2W}^2
\end{array}\right)\left(\begin{array}{c}
\overline{A}_{\mu}^\p\\
\overline{Z}_{\mu}^{\p}
\end{array} \right),\label{e.T2HDMMassMat}
 \eeq
 which admits no massless state for $m_{\overline{A}'}^2>0$.

\begin{figure}
\centering
\includegraphics[trim={8mm 11mm 8mm 8mm},clip,width=0.6\textwidth]{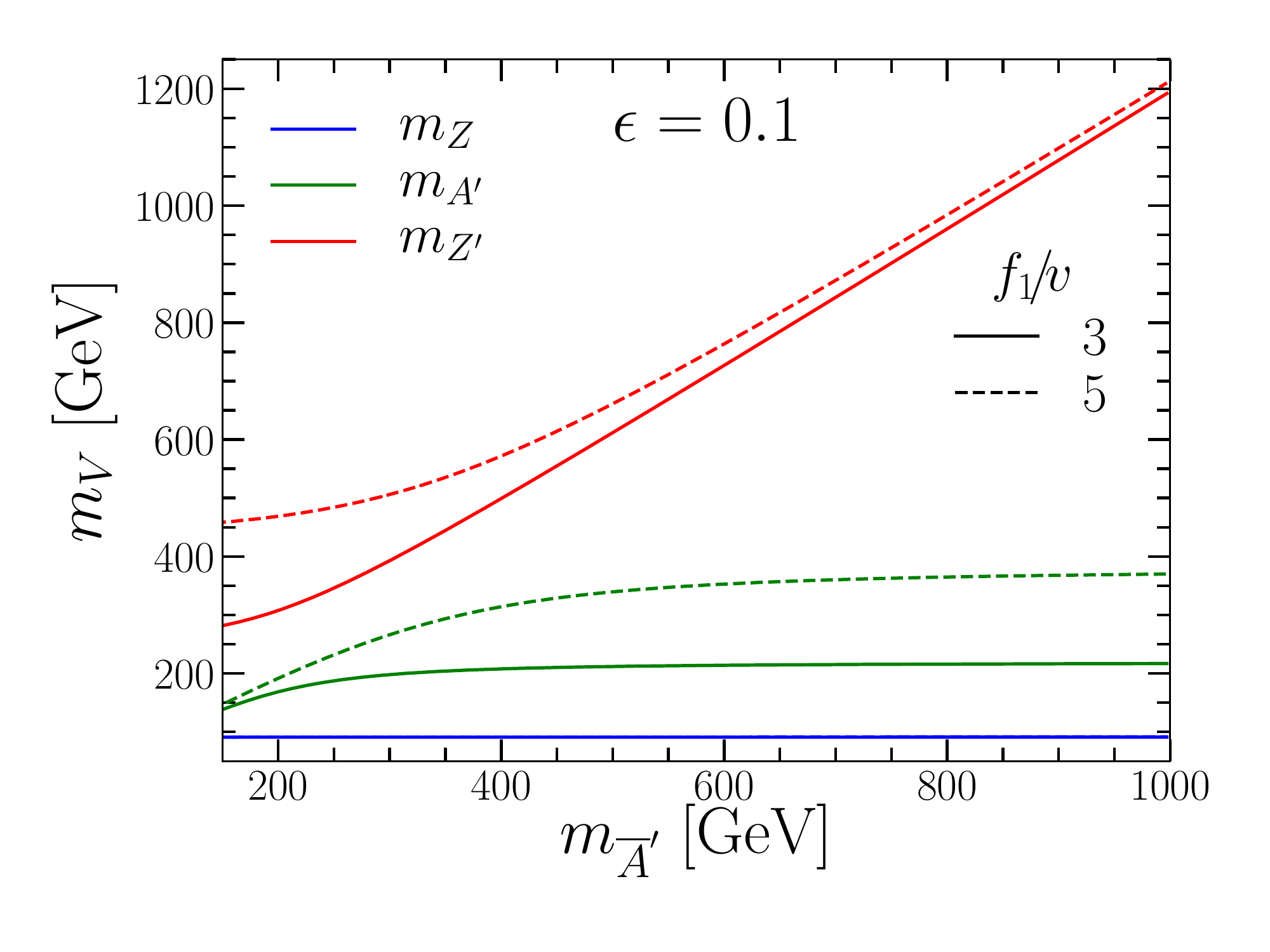}\vspace{-0.25cm}
\caption{The mass eigenvalues for the twin neutral vector bosons as a function of the twin photon mass $m_{\overline{A}'}$ in the T2HDM for $\epsilon=0.1$ and $f_{1}/v=3$ and 5.}
\label{fig:masseigenvalues2}
\end{figure}
From this point, the determination of the physical states and couplings 
in this model proceeds just as in the THPM model, with an example of how the physical masses depend on $m_{\overline{A}^\p} $ given in Fig.~\ref{fig:masseigenvalues2}. The mixing angle that 
diagonalizes the mass matrix in Eq.~(\ref{e.T2HDMMassMat}) is now given 
by
 \begin{equation}
 \sin2\phi= -\frac{m_{\overline{A}'}^2s_{4W}}{\sqrt{\left(m^2_{\overline{Z}^\p}s_{2W}^2+c_{4W}m^2_{\overline{A}^\p} \right)^2+s_{4W}^2m^4_{\overline{A}^\p}}}\,.
 \end{equation}
 In the limit $m^2_{\overline{A}'} \ll m^2_{\overline{Z}^\p}$, the 
mixing angle $\phi$ tends to $-\pi/2$, and the mass eigenstates become, to a 
good approximation, just $\overline{A}_{\mu}^{\p}$ and 
$\overline{Z}_{\mu}^{\p}$, with masses $m_{\overline{A}'}$ and 
$m_{\overline{Z}^\p}$ respectively. In this limit the lighter 
eigenstate, the $A'$, couples more strongly to the visible sector, while 
the couplings of the heavier $Z'$ are suppressed by a relative factor of 
$\tan\theta_W$. In the opposite limit, $m^2_{\overline{A}'} \gg 
m^2_{\overline{Z}^\p}$, the mass eigenvalues are given by $m_{\overline{Z}^\p} 
\sin 2\theta_W$ and $m_{\overline{A}^\p}/\sin 2\theta_W$. In this limit 
it is again the lighter eigenstate, the $A'$, that couples more strongly 
to the visible sector, while the couplings of the heavier $Z'$ are 
suppressed by the same relative factor of $\tan\theta_W$. This is very 
different from the THPM model, in which the couplings of the $A'$ to the 
visible sector vanish in the limit that the $Z'$ is heavy.

\begin{figure}
\centering
\includegraphics[trim={8mm 11mm 8mm 8mm},clip,width=0.49\textwidth]{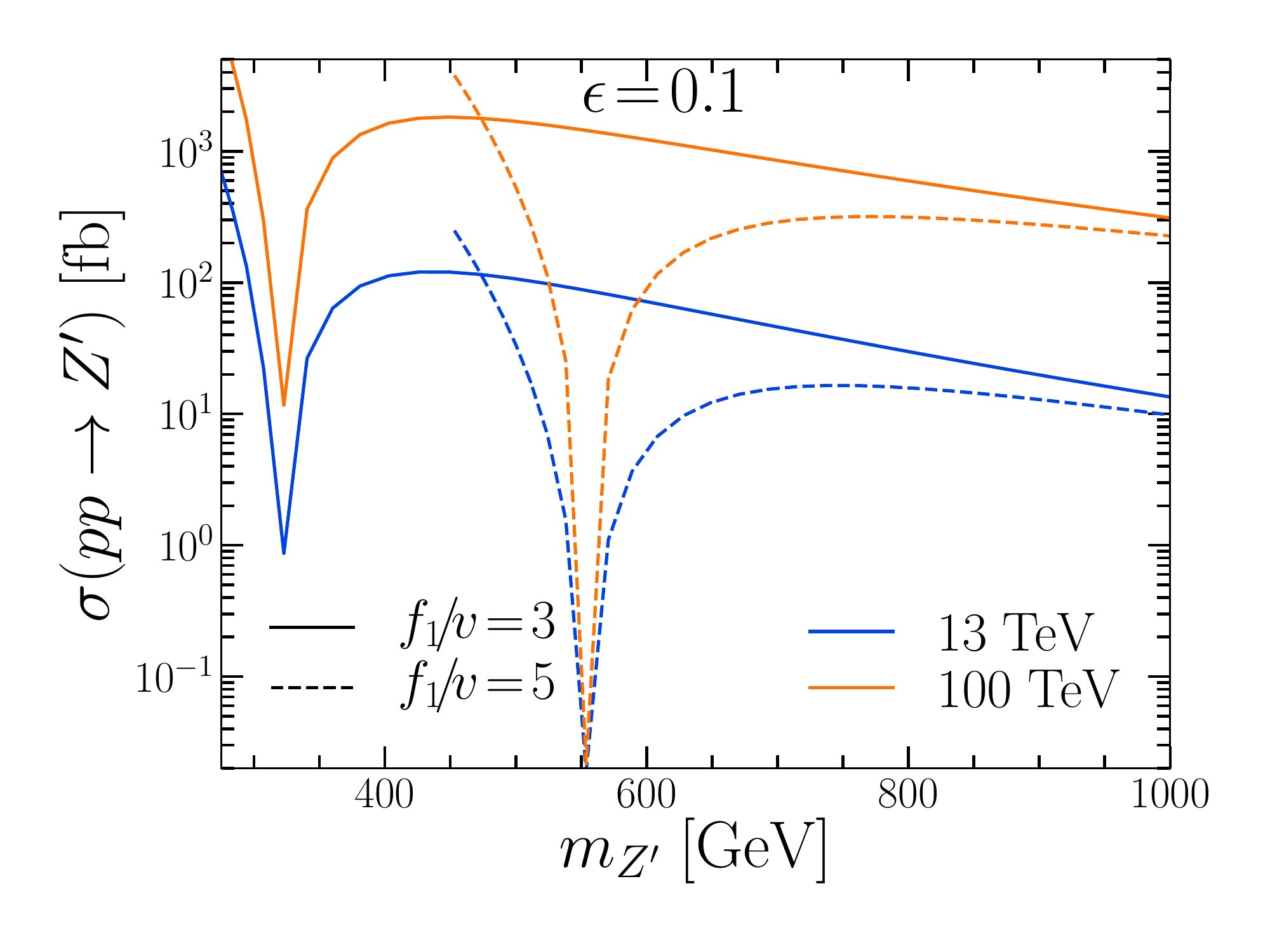}
\includegraphics[trim={8mm 11mm 8mm 8mm},clip,width=0.49\textwidth]{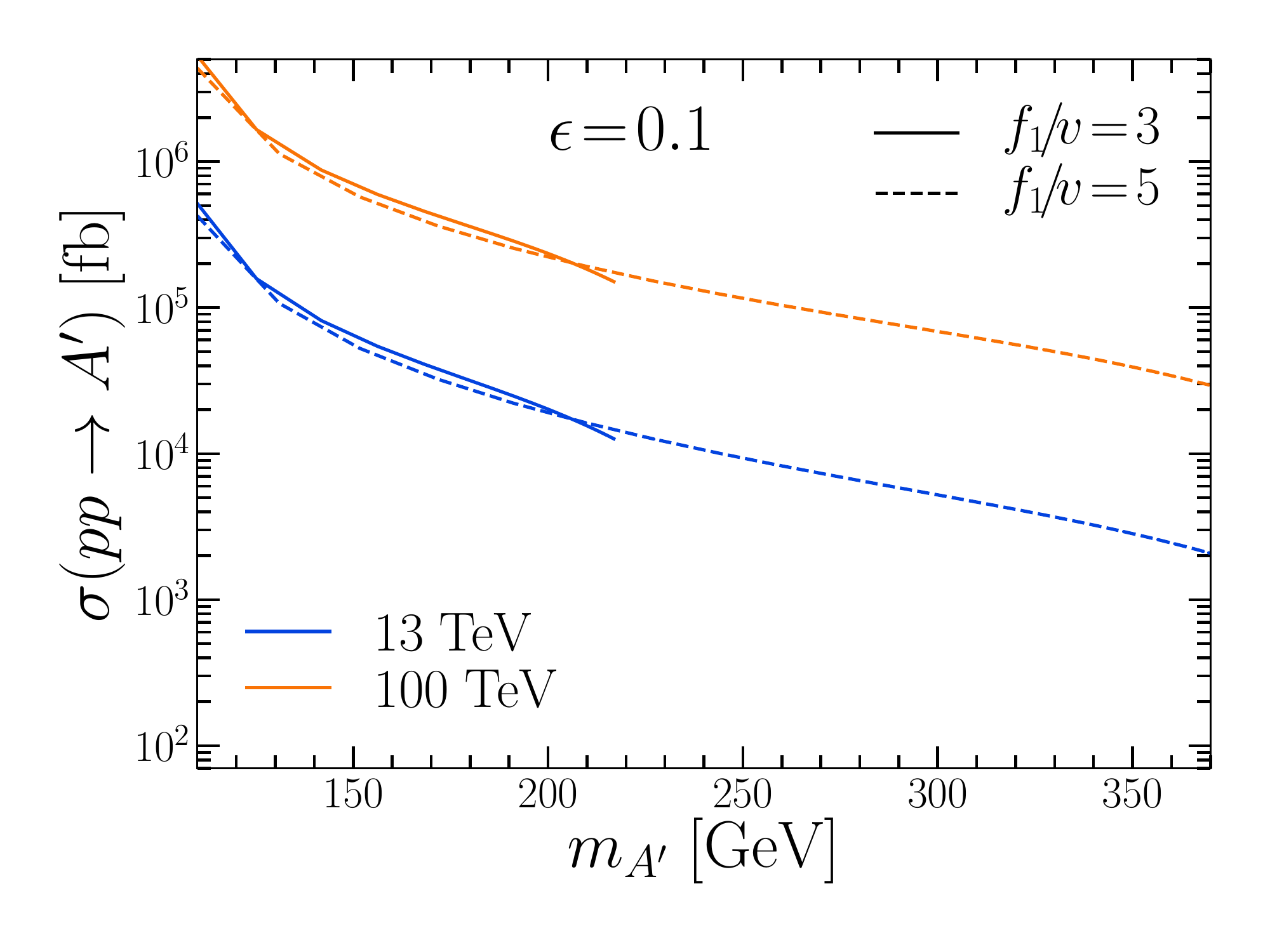}
\caption{Production cross section of the $Z'$ (left) and $A'$ (right) at the LHC and at a future 100~TeV hadron collider in the T2HDM for $\epsilon=0.1$ and $f_1/v=3$ and 5. For $m_{\overline{A}'}=s_{2W}m_{\overline{Z}'}$ the alignment of $Z'$ is orthogonal to twin hypercharge. This closes the portal to the SM sector, reducing production.}
\label{fig:productionxsec2HDM}
\end{figure}

\begin{figure}
\centering
\includegraphics[trim={8mm 8mm 8mm 8mm},clip,width=0.49\textwidth]{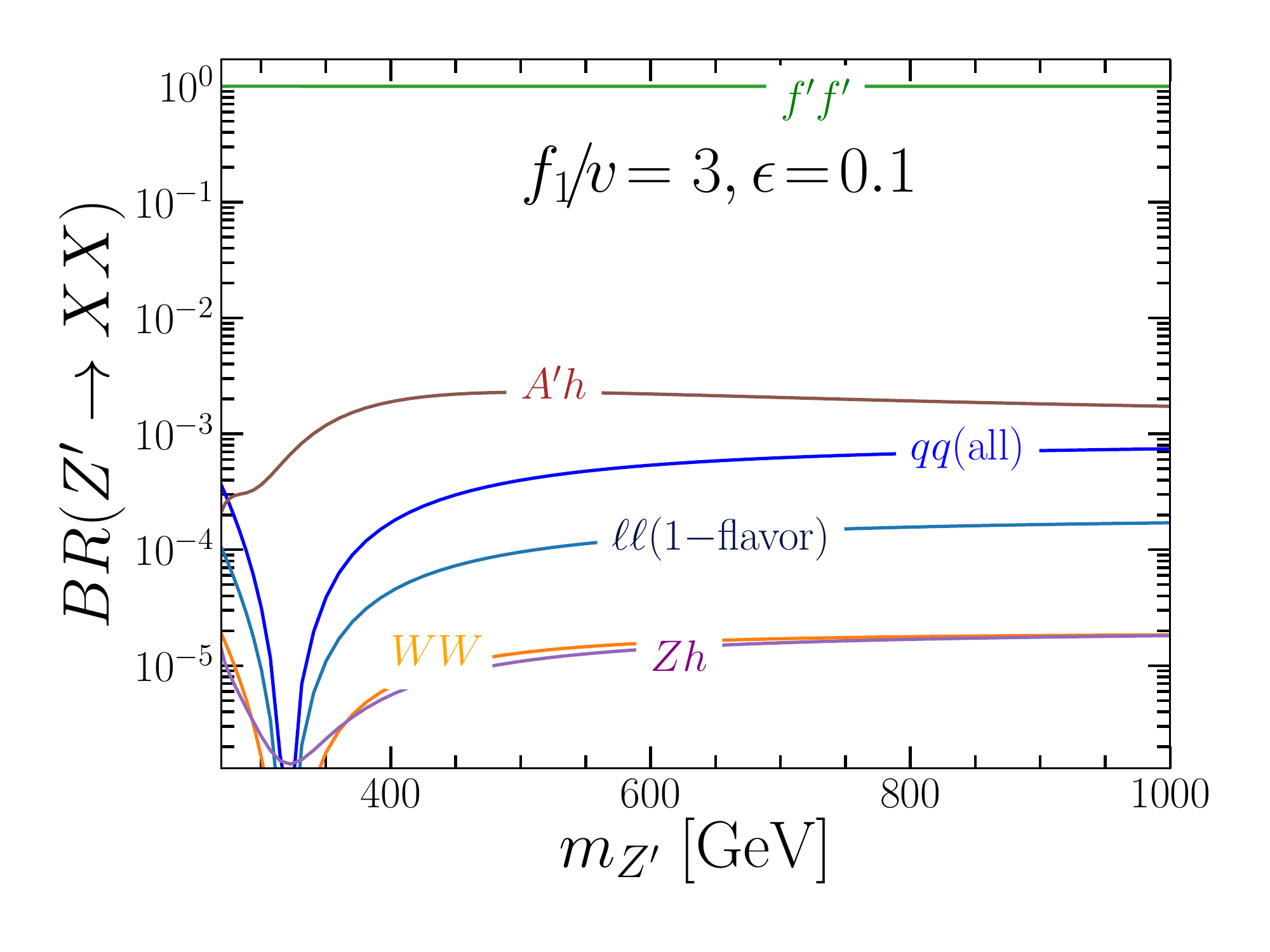}
\includegraphics[trim={8mm 8mm 8mm 8mm},clip,width=0.49\textwidth]{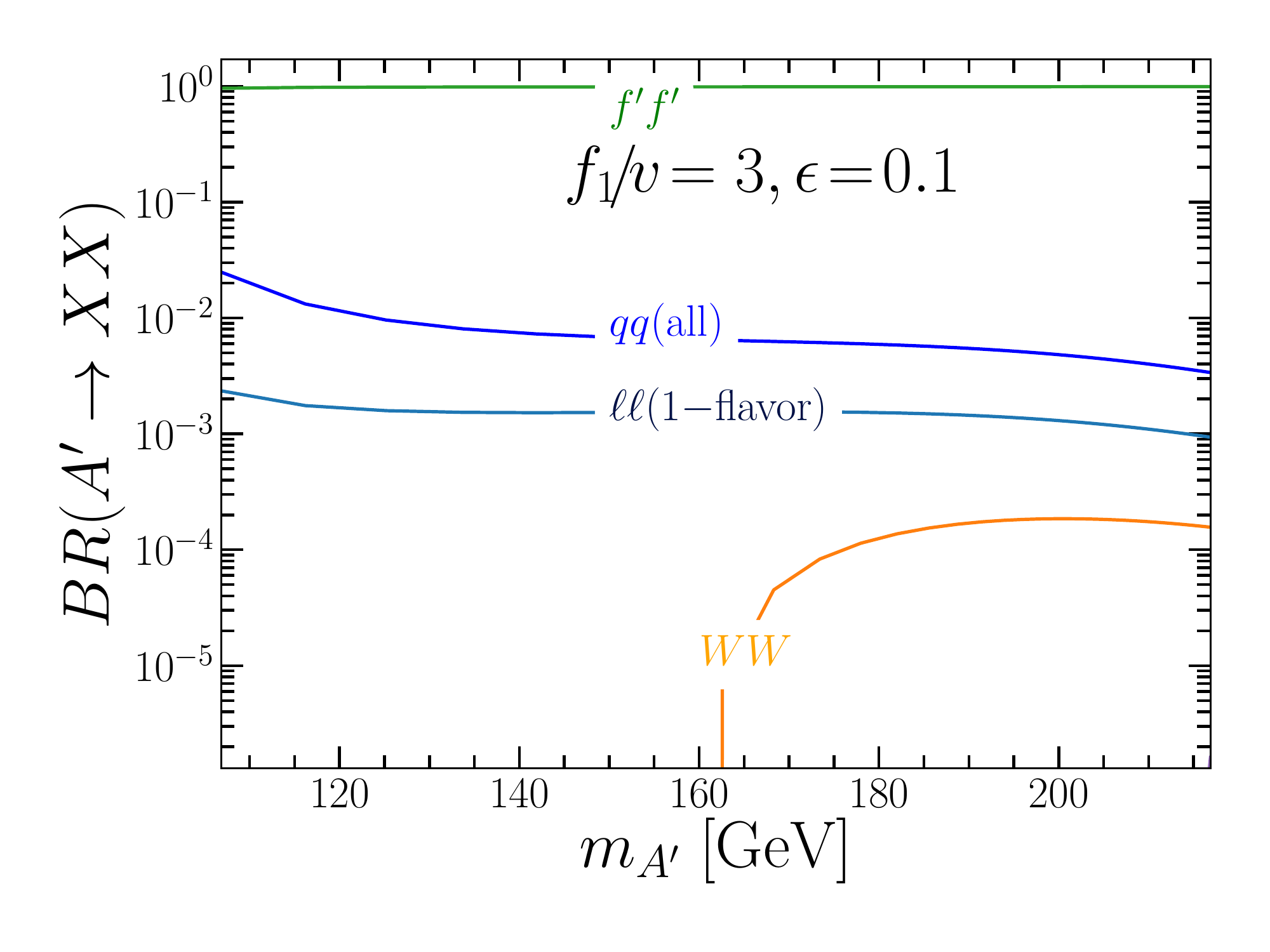}\\
\includegraphics[trim={4mm 5mm 4mm 3mm},clip,width=0.49\textwidth]{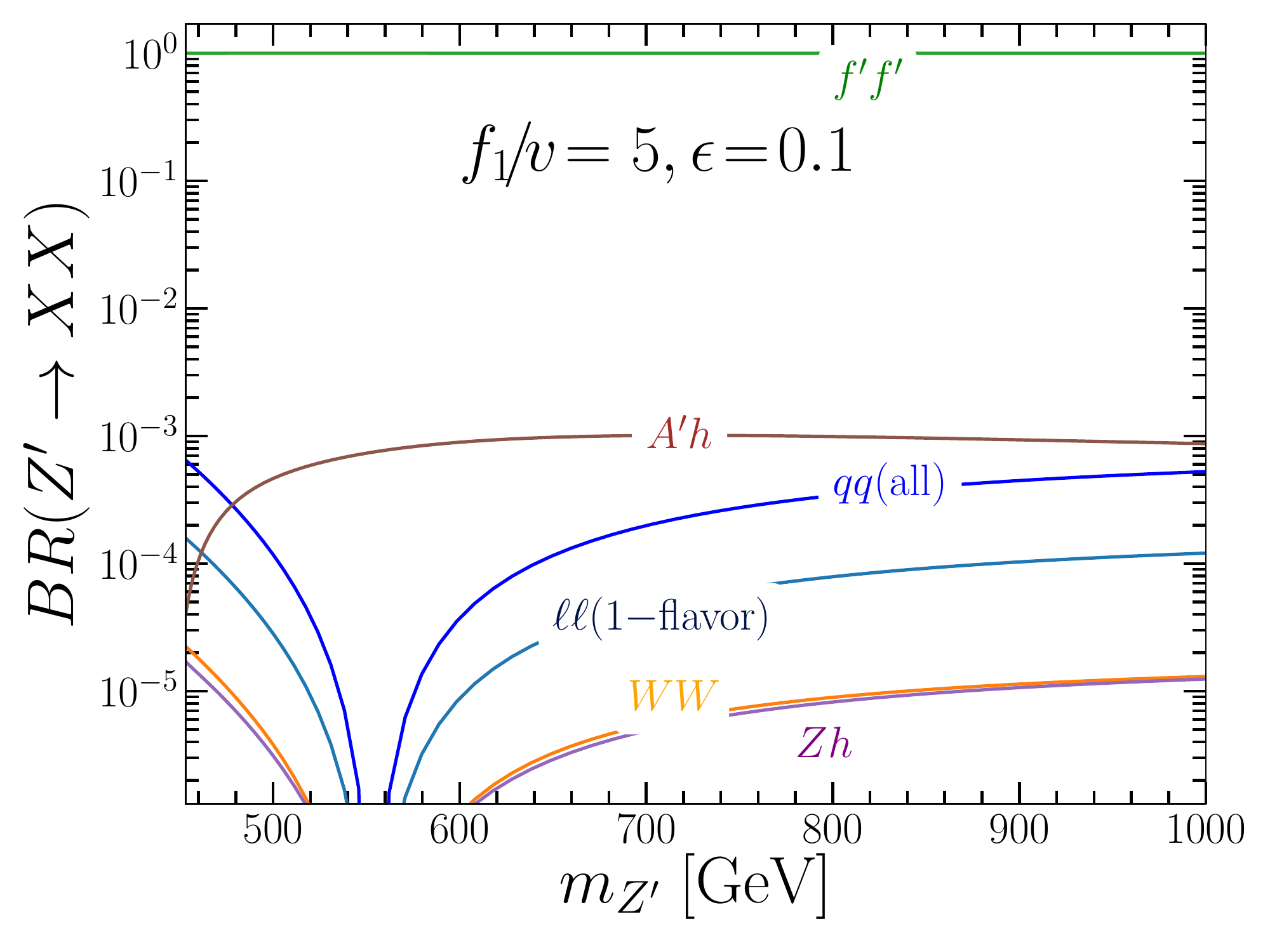}
\includegraphics[trim={3mm 5mm 3mm 3mm},clip,width=0.49\textwidth]{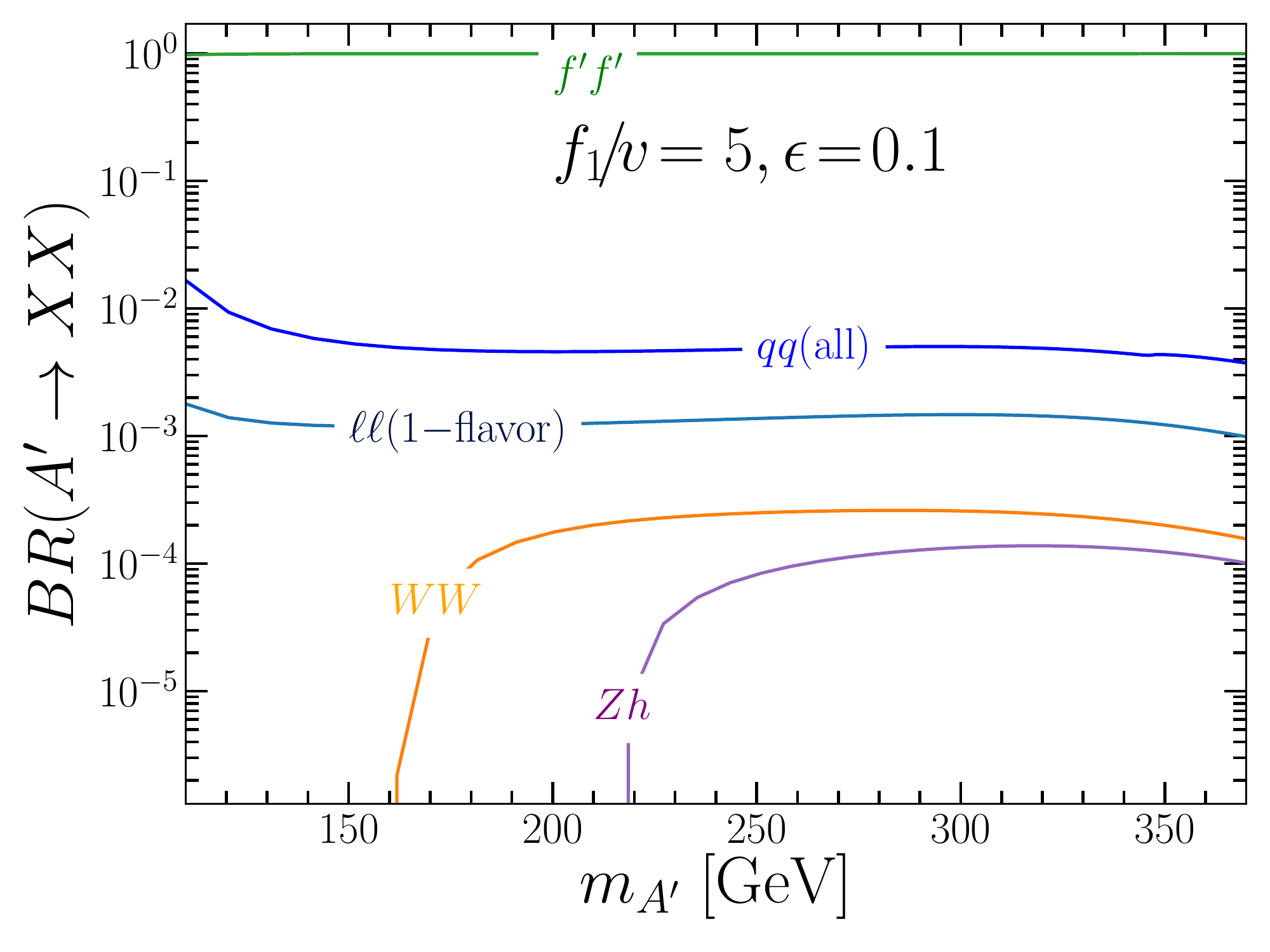}
\caption{Branching ratios of the $Z'$ (left) and $A'$ (right) in the T2HDM for $\epsilon=0.1$ and $f_1/v=3$ and 5. The blue curve is for a single SM lepton flavor while the dark blue curve corresponds to all six quark flavors. For $m_{\overline{A}'}=s_{2W}m_{\overline{Z}'}$ the alignment of $Z'$ is orthogonal to twin hypercharge. This closes the portal to the SM sector, reducing branching to visible states.}
\label{fig:BRs2HDM}
\end{figure}

 At the intermediate value $m_{\overline{A}'}=s_{2W}m_{\overline{Z}'}$, 
corresponding to $\phi=\theta_W-\pi/2$, the $\overline{Z}'$ field 
rotates into $W'^3$, which means it is orthogonal to $B'$. As we see in 
Figs.~\ref{fig:productionxsec2HDM} and~\ref{fig:BRs2HDM}, this causes 
the $Z'$ to completely decouple from the visible sector, suppressing the 
production cross section and branching into visible states. At the same 
point, of course, the $A'$ is perfectly aligned with twin hypercharge, so its coupling to the visible sector is enhanced.

To analyze the experimental signals, we transform to a basis in which 
the mass and kinetic terms of the $A'$ and $Z'$, as well as those of the 
SM photon and $Z$, are diagonal. As in the THPM case, this is 
accomplished by first performing a shift of the SM hypercharge gauge 
boson $B$ and then rotating into the mass basis, as discussed in 
Appendix~\ref{a.Diag}. Once the kinetic and mass terms of the vector 
bosons have been diagonalized, it is straightforward to calculate their 
couplings to fermions. The details can be found in 
Appendix~\ref{a.VFcouplings}. In Figs.~\ref{fig:productionxsec2HDM} 
and~\ref{fig:BRs2HDM} we plot the production cross sections and 
branching fractions, respectively, of the $A'$ and $Z'$ in the T2HDM. We 
see that the production cross section for the $A'$ is always larger than 
for the $Z'$. The branching fraction of the $A'$ to SM final states is 
also greater than for the $Z'$. These features, along with its lighter 
mass, enhance the prospects for $A'$ discovery while inhibiting $Z'$ 
discovery. We also see a striking feature when 
$m_{\overline{A}'}=s_{2W}m_{\overline{Z}'}$, where the $Z'$ is 
orthogonal to twin hypercharge and its couplings to SM states vanish.


\section{Existing Constraints\label{sec:constraints}}

\begin{figure}
\centering
\includegraphics[trim={9mm 8mm 9mm 9mm},clip,width=0.49\textwidth]{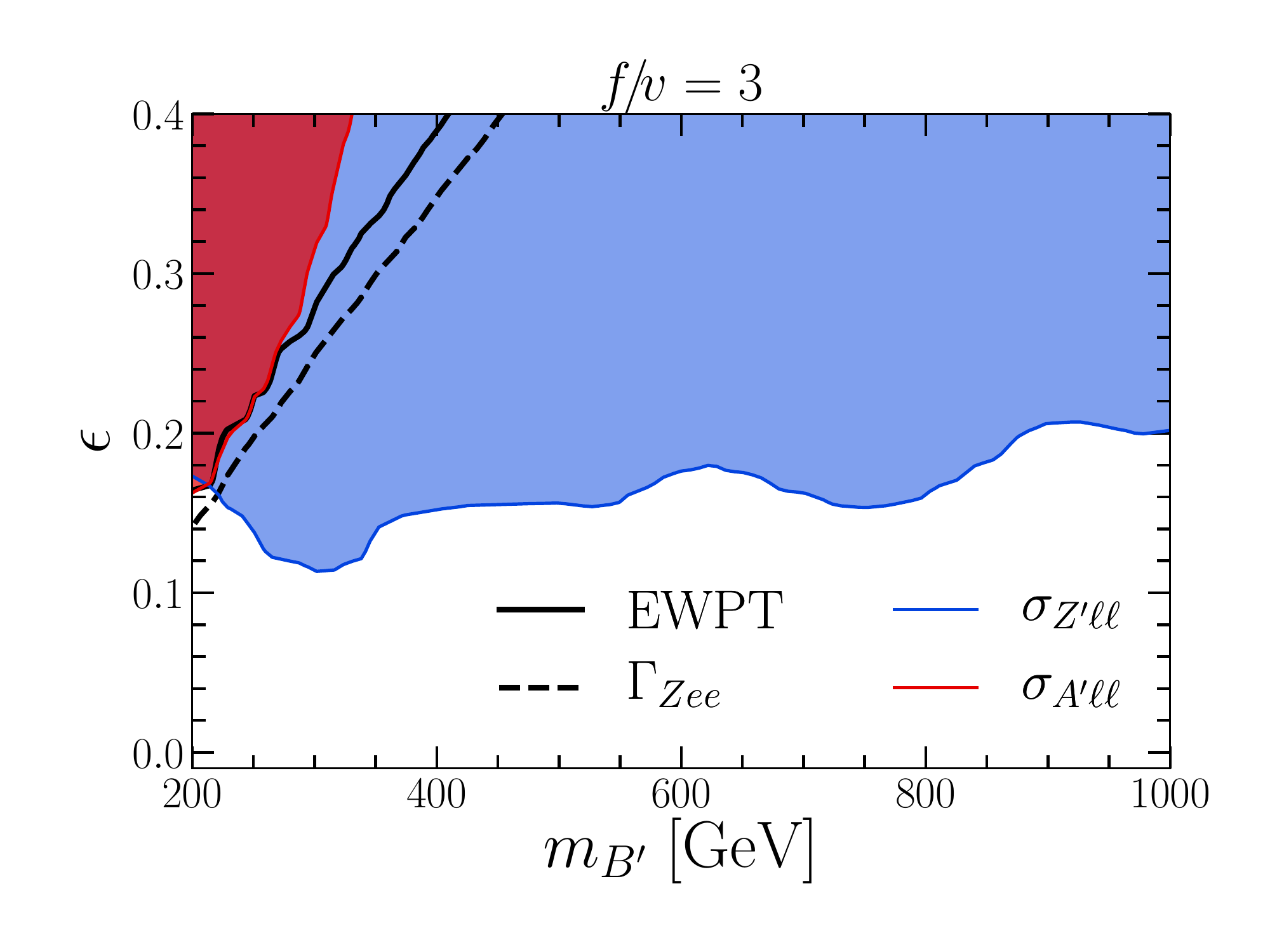}
\includegraphics[trim= 9mm 7mm 9mm 6mm,clip,width=0.49\textwidth]{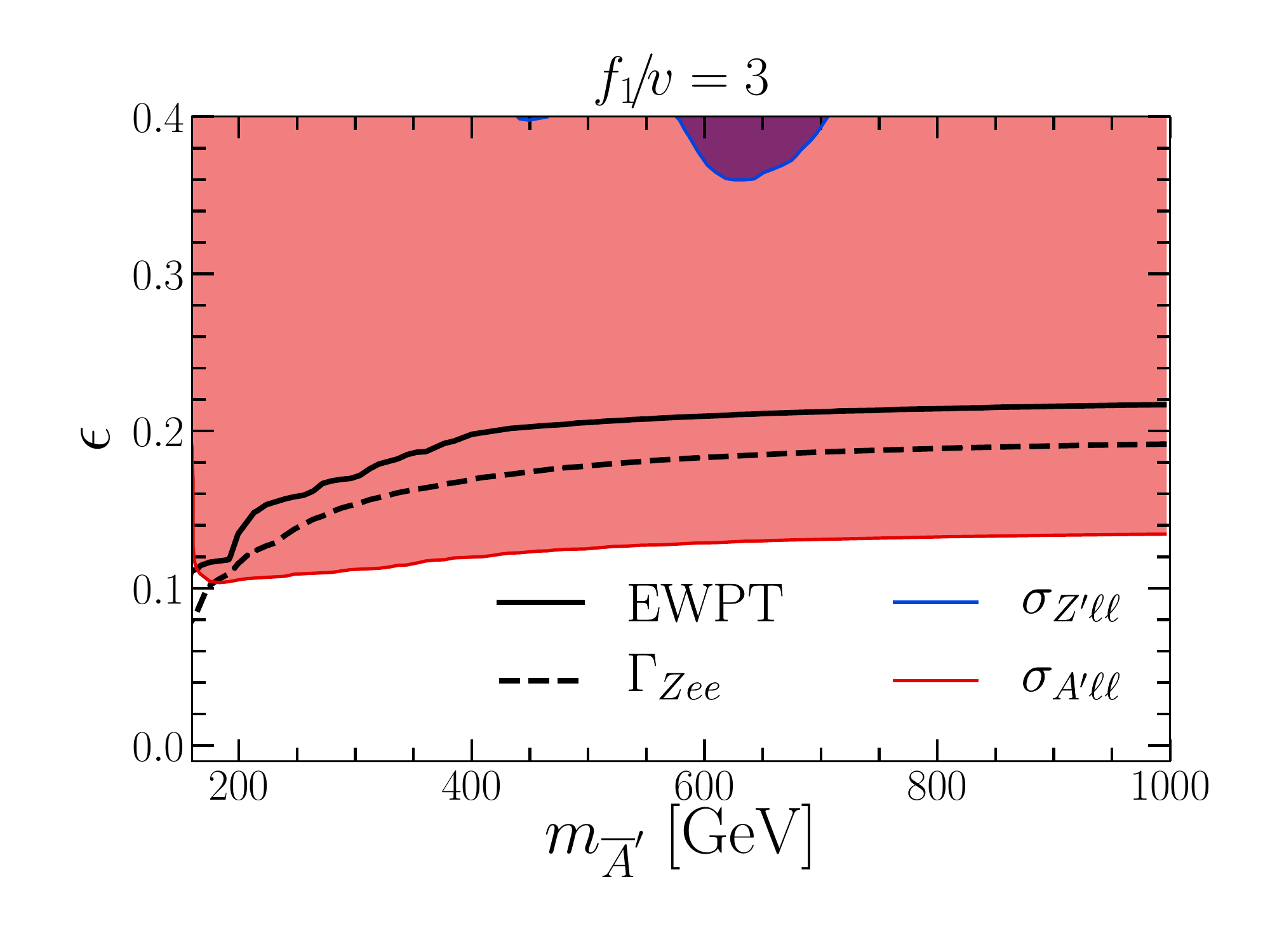}\\
\includegraphics[trim={9mm 10mm 9mm 8mm},clip,width=0.49\textwidth]{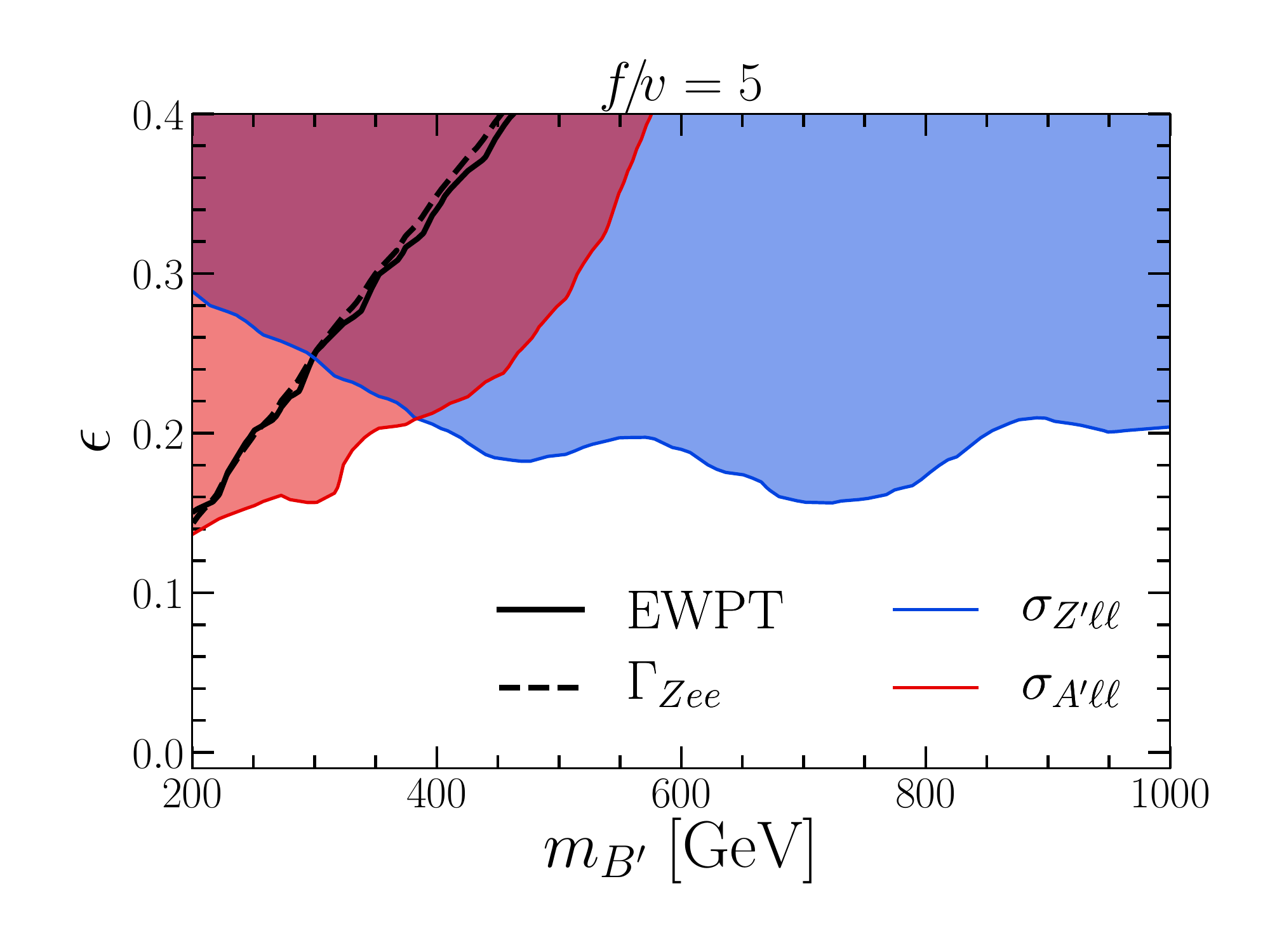}
\includegraphics[trim={9mm 10mm 9mm 8mm},clip,width=0.49\textwidth]{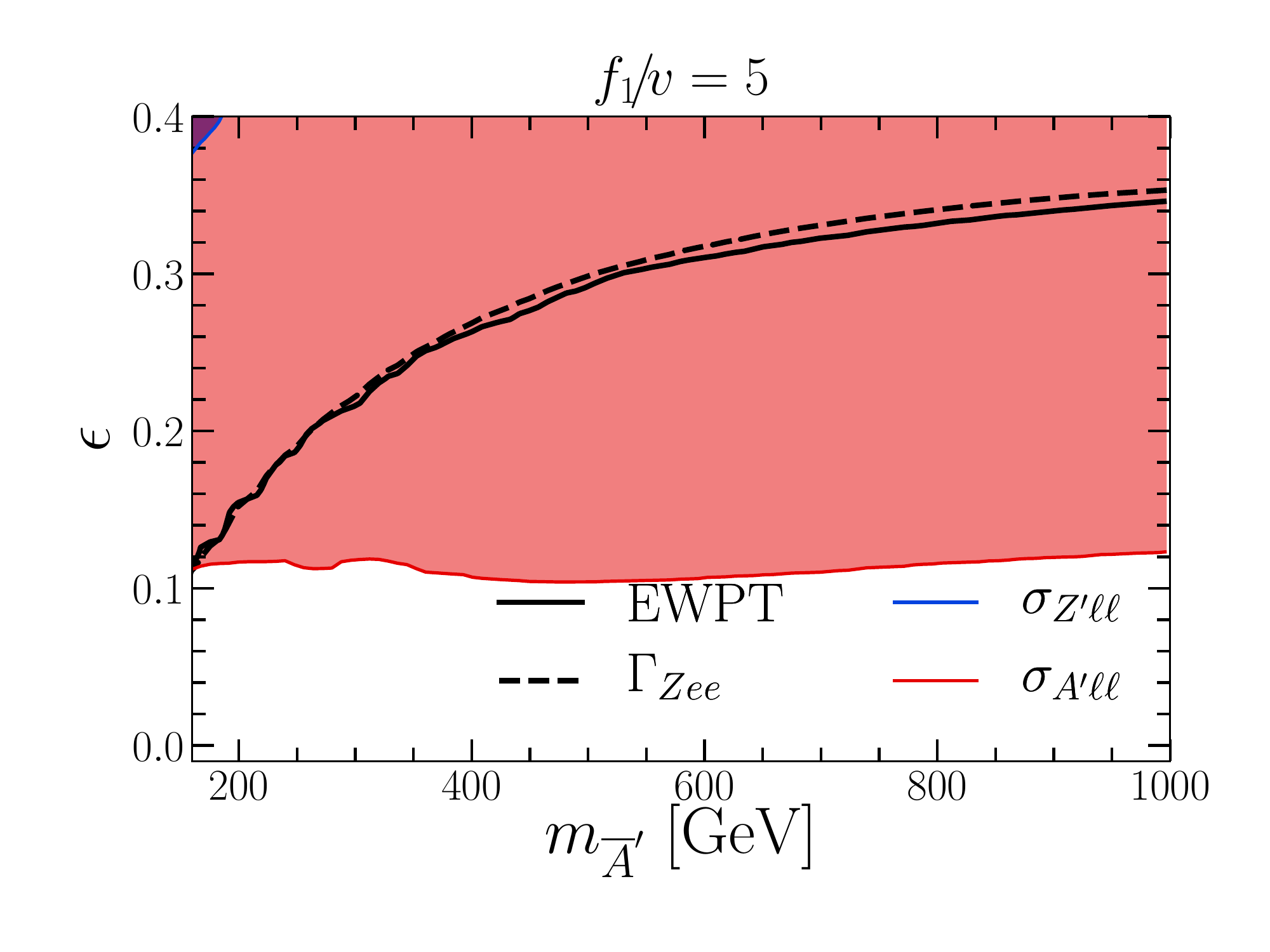}
\caption{Compilation of constraints on the MTH model in terms of $\epsilon$ and either $m_{B^\p}$ (THPM model on the left) or $m_{\overline{A}'}$ (T2HDM on the right) with $f_{(1)}/v=3$(5) in the top (bottom) row. Indirect constraints are shown in black. The 13 TeV LHC dilepton resonance search bounds on the $A'$ ($Z'$) are shown in red (blue). }
\label{fig:constraints}
\end{figure}

In this section we determine the constraints on the vector boson sector 
of the MTH model from direct searches and from precision 
electroweak observables. These are displayed in 
Fig.~\ref{fig:constraints}, the left plot showing the THPM model and the 
right plot the T2HDM. The top (bottom) row is for the benchmark of 
$f_{(1)}/v = 3(5)$.

\begin{itemize}
 \item LEP analyses place strong limits on any new physics contributions that affect the properties of the $Z$ boson. Quantities that can be determined directly from measurements at the $Z$-pole include the total width of the $Z$ boson, as well as the cross sections into various SM final states through the $Z$ resonance. Given these measurements, theory can be employed to fit other parameters in a straightforward way. In particular, theory relates the resonant dilepton production cross section and the total width to the leptonic width of the $Z$ in a model-independent way. This can therefore be used to constrain models of new physics. A different combination of the direct $Z$-pole measurements allows the invisible width of the $Z$ boson to be determined, again in a model-independent way, which provides a powerful independent constraint on new physics. We therefore begin our study of existing constraints on the MTH model with these two LEP $Z$-pole measurements.

The measured invisible width of the $Z$-boson is $\Gamma_\text{Inv}^\text{LEP}=499.0\pm 1.5\, 
\text{MeV}$~\cite{Patrignani:2016xqp}. The predicted value for the SM 
contribution is $\Gamma_\text{Inv}^\text{SM}=501.3\pm 0.6$ 
MeV~\cite{Carena:2003aj}. Therefore, the preferred central value and 
associated uncertainty of any potential new physics contribution are 
given by
 \beq
\Delta\Gamma_\text{Inv}=-2.2\pm 1.6 \mev. \label{difference}
 \eeq
 In our models, there are two contributions to 
$\Delta\Gamma_\text{Inv}$, a {\it reduction} due to the change in the 
couplings to the neutrinos, and a strictly positive contribution due to 
$Z$ decays to kinematically accessible twin states. We find  that 
the reduction in the SM width dominates, so the MTH invisible $Z$ 
width is generically smaller than the SM prediction, bringing it closer to the LEP 
measurement. Consequently, this does not significantly constrain either 
model. 

The modification of the $Z$-boson coupling to electrons in the MTH model, as well as to the twin sector states also constrains the parameter space. The LEP bound on the partial width of $Z\to e^{+} e^{-}$ is given by~\cite{Patrignani:2016xqp}
\beq
\Gamma\left(Z \to e^+ e^-\right) = 83.91\,\pm\, 0.12 ~\text{MeV}.
\eeq
We require this partial width to be within the 2$\sigma$ range from 83.67 MeV to 84.15 MeV in our model, which excludes the regions above the black dashed lines in Fig.~\ref{fig:constraints}.

\item 
 Electroweak precision data as encoded in the oblique 
parameters~\cite{Peskin:1990zt}, in particular the $T$ parameter, also 
constrain the relevant parameter space. The mixing among the neutral 
vector bosons alters the relation between the $Z$ boson mass and the $W$ 
boson mass in the SM. The leading order correction is given in 
Eq.~\eqref{e.massesZ}. Using the notation of 
ref.~\cite{Appelquist:2002mw} we have
 \beq
\frac{\alpha T}{1+\alpha T}=1-\frac{m_Z^2}{m_{Z_0}^2}.\label{e.leadingT}
\eeq
 Here $m_{Z_0}$ is the $Z$ boson mass in the SM, and $\alpha$ is the 
fine structure constant evaluated at $m_{Z_0}$.

\hspace{18pt}While this effect is the largest, there are other 
contributions to both the $S$ and $T$ parameters from the reduction of 
the Higgs coupling to SM states. These effects have been calculated in 
the general case~\cite{Falkowski:2013dza}. Applied to our models, we obtain
 \begin{equation}
T\approx-\frac{3v^2}{8\pi c_W^2 f^2_{(1)}}\ln\frac{\Lambda_\text{UV}}{m_{Z_0}}, \ \ \ \ S\approx\frac{v^2}{6\pi f^2_{(1)}}\ln\frac{\Lambda_\text{UV}}{m_{Z_0}}.
\label{S&T}
 \end{equation}
 While this contribution to $T$ is smaller than Eq.~\eqref{e.leadingT}, 
it also has the opposite sign, reducing the deviation. We use the 
current bound on the $T$-parameter leaving $U$ a free 
parameter~\cite{Baak:2014ora}. For $\Lambda_\text{UV}=5$ TeV the 
contribution to $S$ in Eq.~(\ref{S&T}) varies from about 0.02 to 0.008, 
while the contribution to $T$ varies from $-0.06$ to $-0.02$, as $f/v$ 
changes from 3 to 5. In this case, the 95\% exclusion contours require 
$T<0.14$ and 0.13 respectively. The excluded region is shown as the area 
above the solid black line in Fig.~\ref{fig:constraints}.

\hspace{18pt} We see from the figure that both the $Z\to e^+e^-$ partial 
width and the $T$-parameter bounds are extremely restrictive in the 
T2HDM case. This occurs because, in the T2HDM, the lighter state 
continues to have sizable couplings to the SM even as $m_{\overline{A}'}$ 
gets large, so the bound remains almost unchanged even as the $Z'$ 
becomes heavy. In contrast, in the THPM model, the lighter $A'$ 
decouples from the SM sector in the limit that the mass of the $Z'$ is 
large, so the bounds on this scenario fall off as $m_{Z'}$ 
increases.

\item 
 Direct searches at the LHC place strong bounds on dilepton resonances 
produced from a $\bar{q}$-$q$ initial state. We use the ATLAS 
limits~\cite{Aaboud:2017buh,Aad:2019fac}, and employ
{\sc{MadGraph5}}~\cite{Alwall:2014hca} to simulate backgrounds. The $A'$ 
and $Z'$ production and decay at the LHC are simulated using the MSTW 
PDFs~\cite{Martin:2009iq}. Our analysis is performed at the parton level 
since we are only dealing with leptons in the final state. In 
Fig.~\ref{fig:constraints} we show the 95\% exclusion region from the 
resonant $Z'$ (blue shaded) and $A'$ (red shaded) production with 
subsequent decays to electrons and muons as a function of $m_{B'}$ 
(left) or $m_{\overline{A}'}$ (right) and $\epsilon$ and 
$f_{(1)}/v=3(5)$ on the top (bottom) row.

\hspace{18pt} As we saw in Fig.~\ref{fig:masseigenvalues}, the $A'$ mass 
in the THPM model asymptotes to $(v'/v) m_{W}$ as $m_{B'}$ gets large, 
with similar behavior in the T2HDM. However, in the THPM case both the 
production cross section (see Fig.~\ref{fig:productionxsec}) and 
dilepton branching ratio of the $A'$ (see Fig.~\ref{fig:BRs}) fall off 
rapidly in this region, where $A'$ decouples from the SM. Therefore, the 
$A'$ is unconstrained for larger values of $m_{B'}$. Conversely, in the 
T2HDM the production (see Fig.~\ref{fig:productionxsec2HDM}) and 
dilepton branching fraction of the $A'$ (see Fig.~\ref{fig:BRs2HDM}) 
remain sizable as $m_{\overline{A}'}$ increases, making the bound from 
$A'$ searches strong over the entire mass range of interest. In the THPM 
model the $Z'$ limits persist, dominating the limits at higher $m_{B'}$. 
In the T2HDM, however, the $Z'$ decouples from the SM at 
$m_{\overline{A}'}=s_{2W}m_{\overline{Z}'}$, causing the bound to vanish 
at this point. Even away from this point, the couplings of the $Z'$ to 
the SM are, in general, smaller than those of the $A'$, so the bounds 
on the T2HDM are dominated by the limits on the $A'$.

\end{itemize}

\section{Discovery Prospects \label{s.discovery}}

\begin{figure}
\centering
\includegraphics[trim={9mm 8mm 9mm 9mm},clip,width=0.49\textwidth]{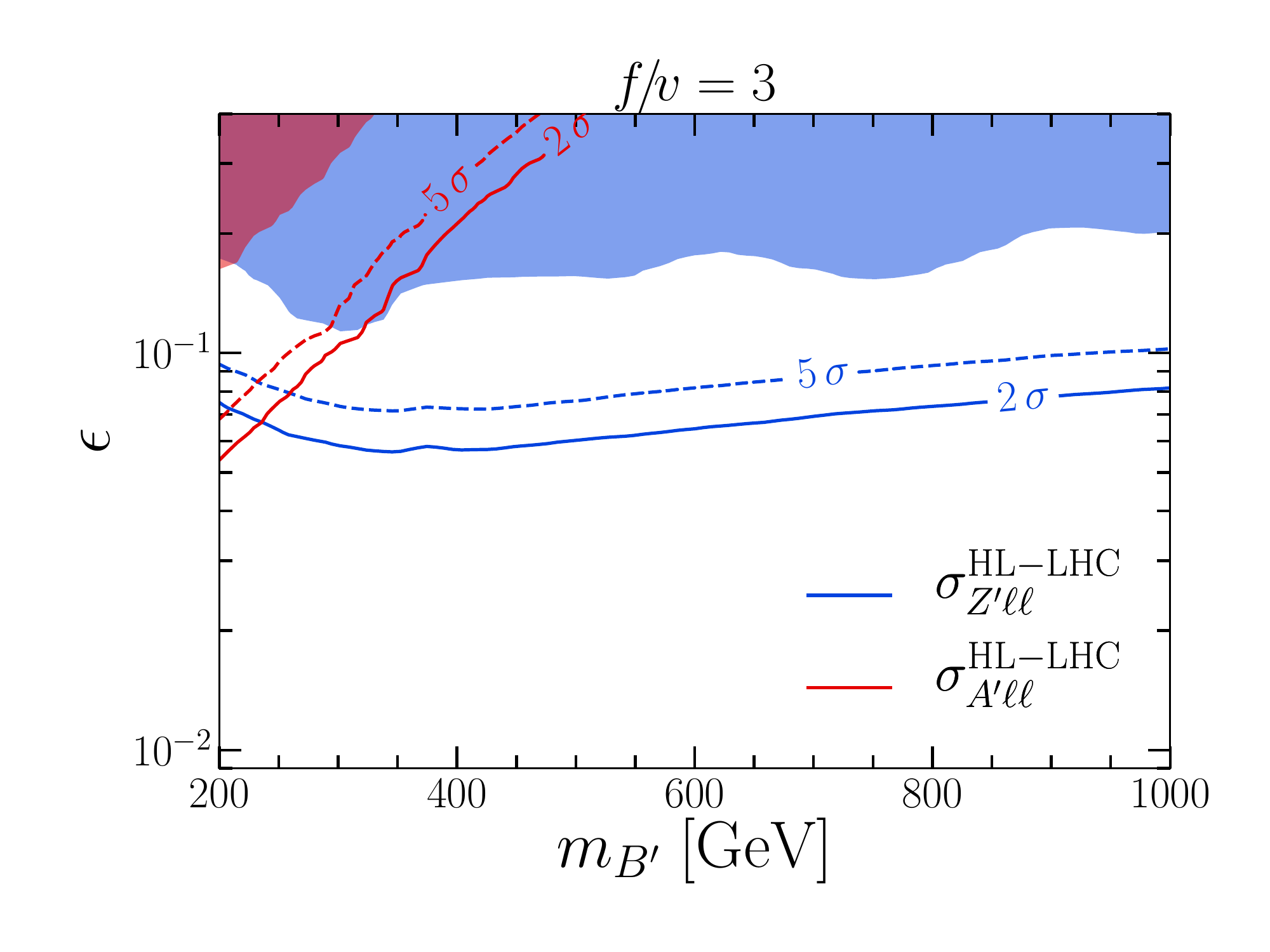}
\includegraphics[trim={9mm 8mm 9mm 9mm},clip,width=0.49\textwidth]{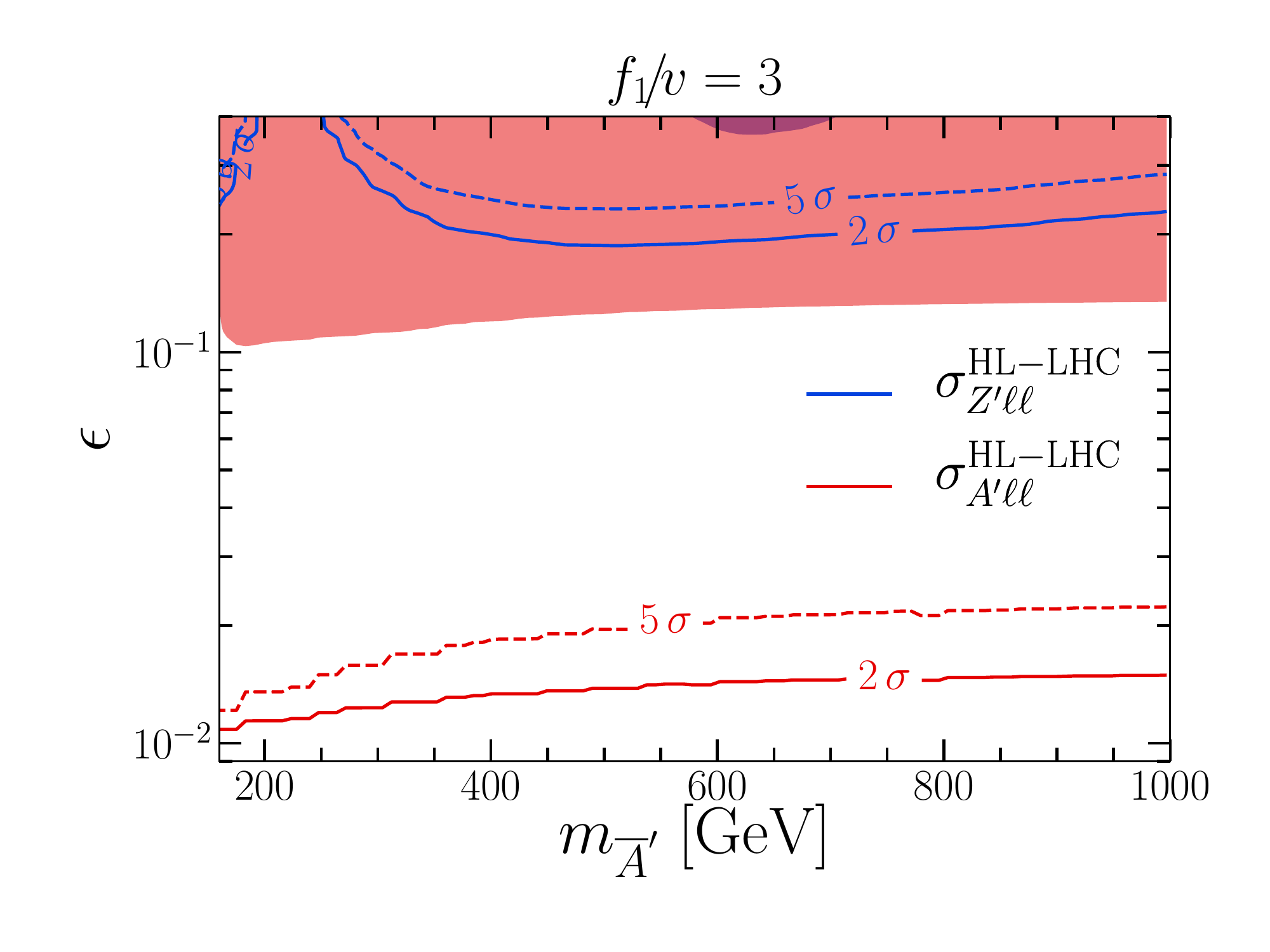}\\
\includegraphics[trim={9mm 8mm 9mm 9mm},clip,width=0.49\textwidth]{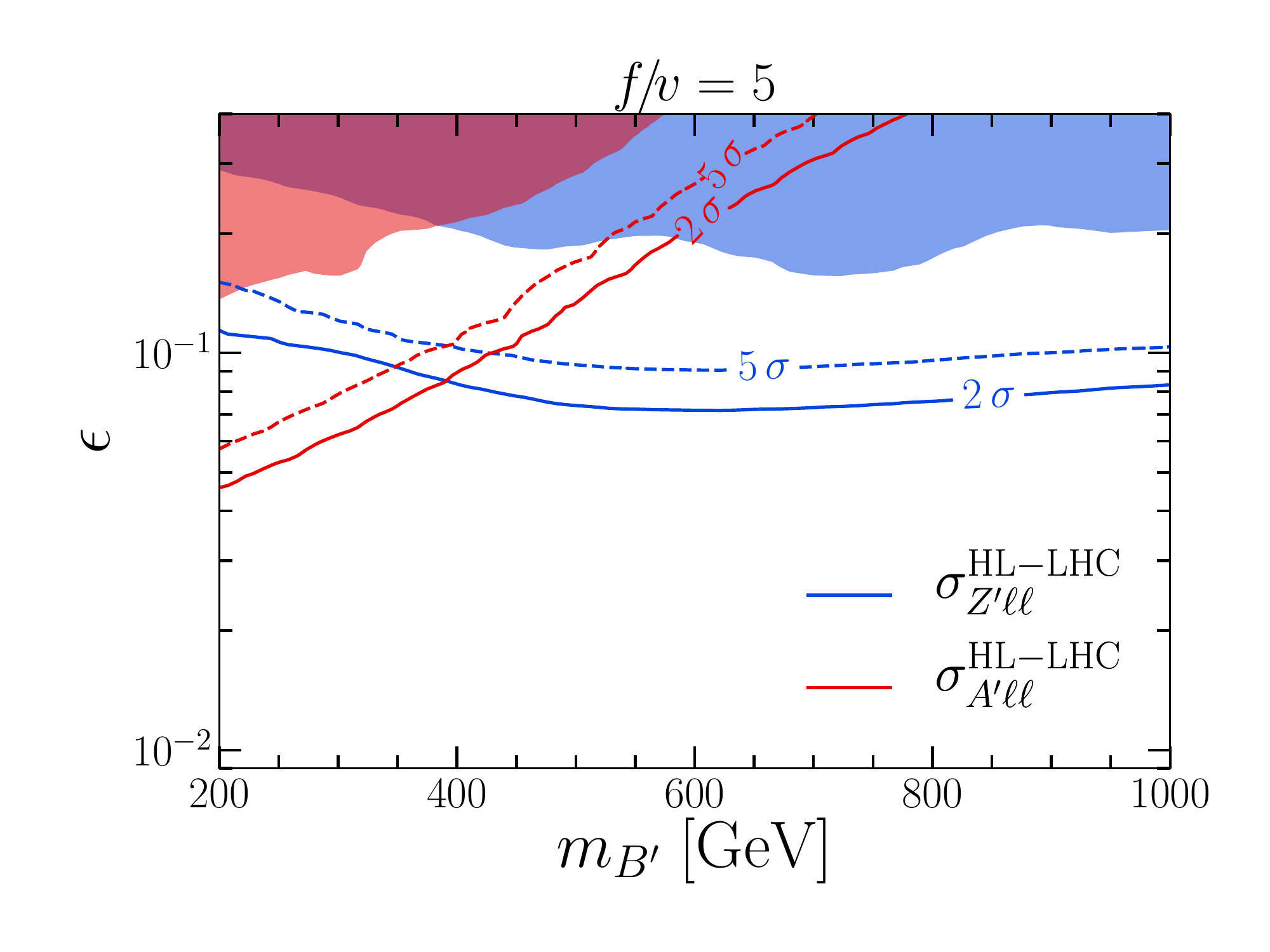}
\includegraphics[trim={9mm 8mm 9mm 9mm},clip,width=0.49\textwidth]{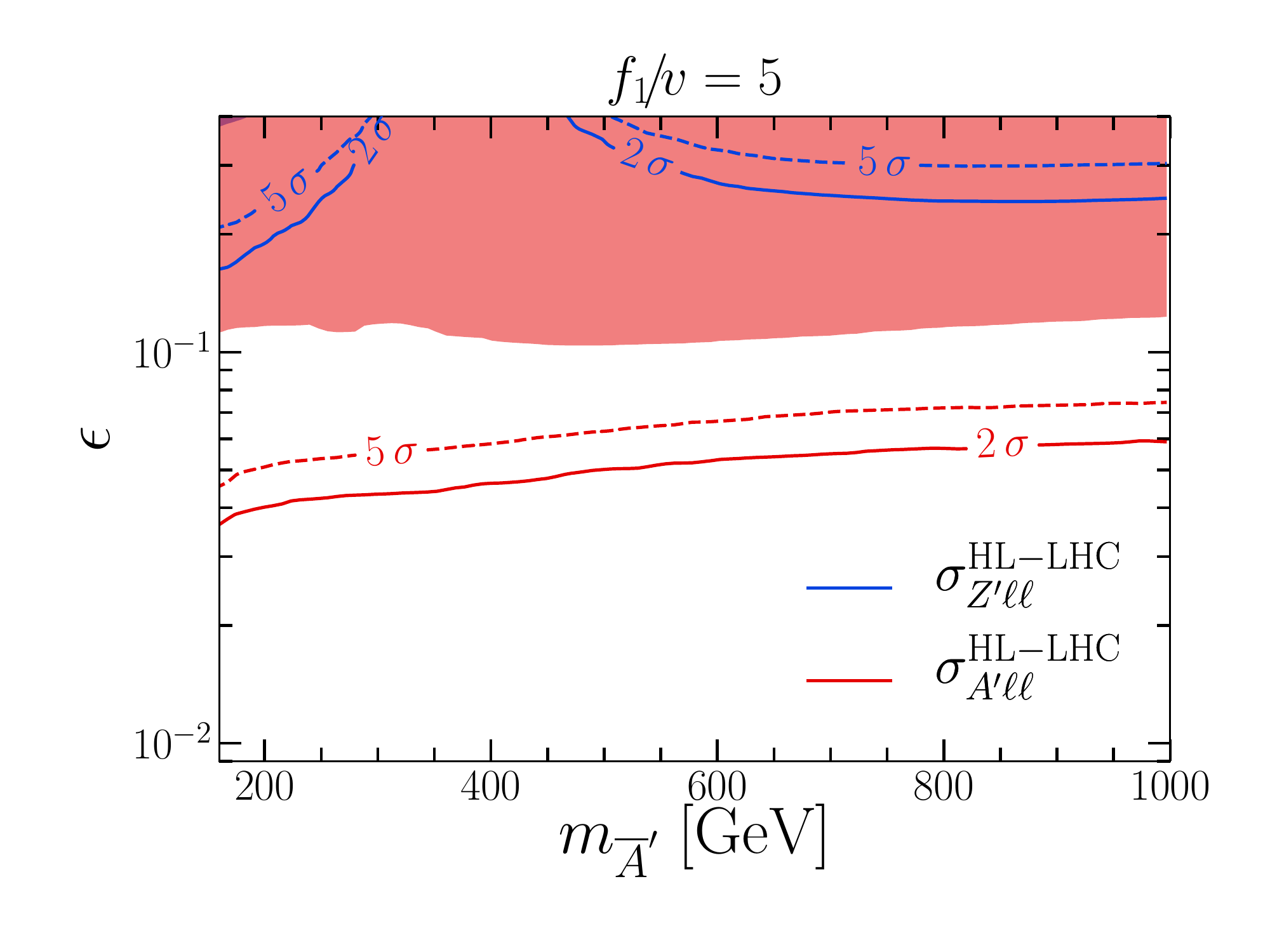}
\caption{Projected $\sigma$ contours of the $Z'$ (blue) and $A'$ (red) at the HL-LHC for a luminosity of 3000~fb$^{-1}$ for $f_{(1)}/v=3$(5) on the top (bottom) row. The THPM model is on the left and the T2HDM is on the right. The regions excluded by current LHC searches for the twin bosons are shaded in the corresponding color.}
\label{fig:futuresensitivityHLLHC}
\end{figure}

\begin{figure}
\centering
\includegraphics[trim={9mm 8mm 9mm 9mm},clip,width=0.49\textwidth]{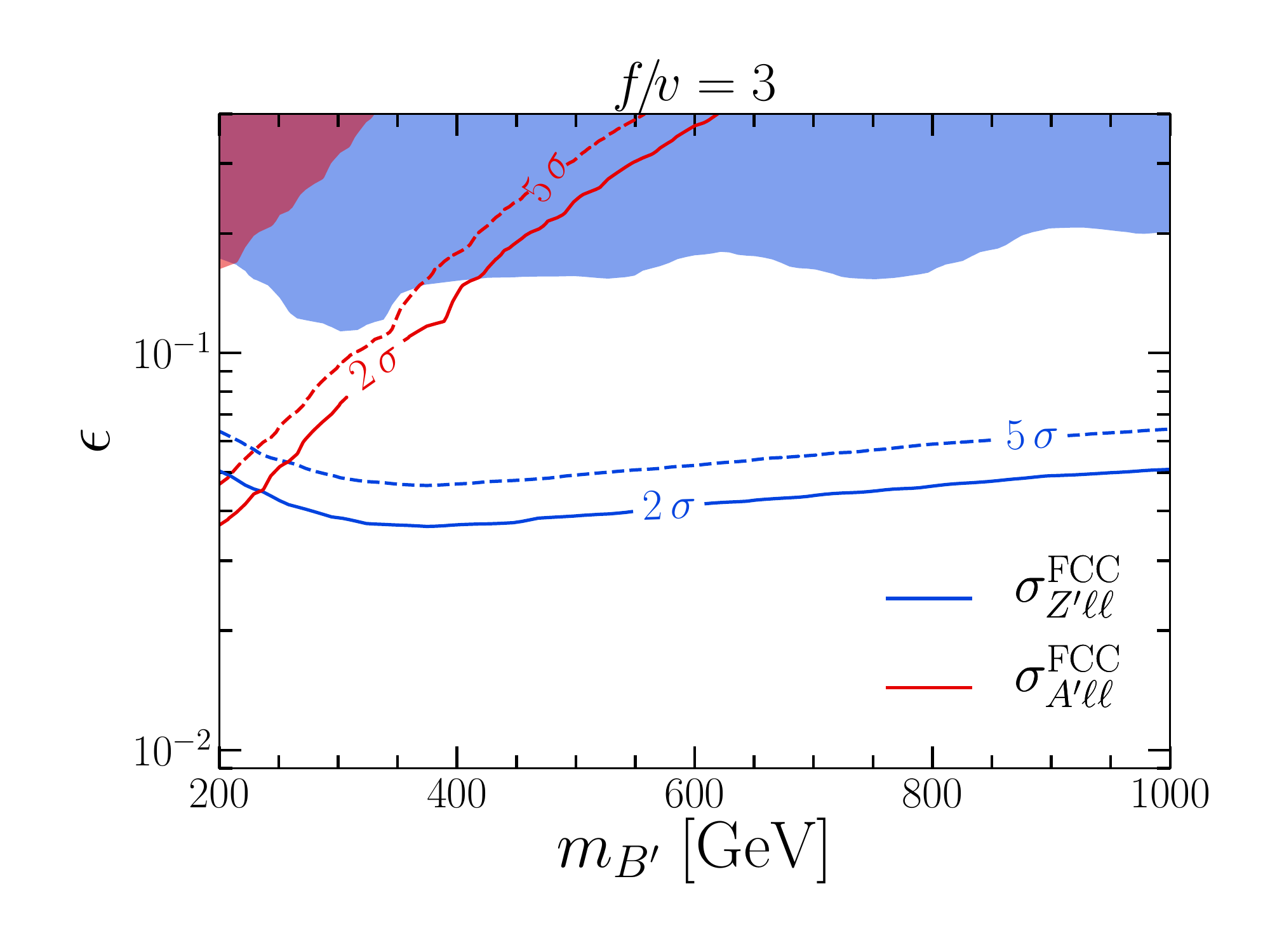}
\includegraphics[trim={9mm 8mm 9mm 9mm},clip,width=0.49\textwidth]{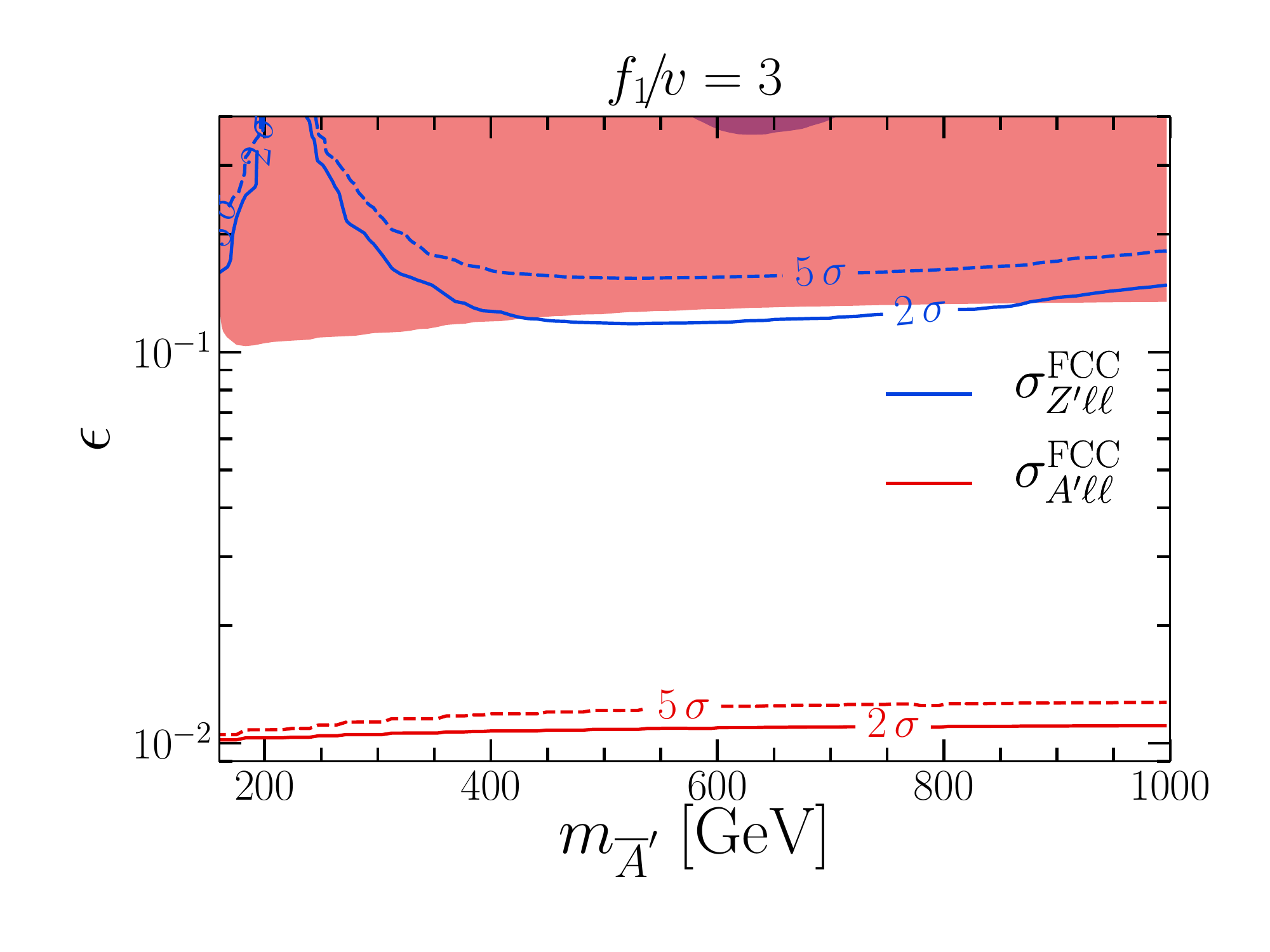}\\
\includegraphics[trim={9mm 8mm 9mm 9mm},clip,width=0.49\textwidth]{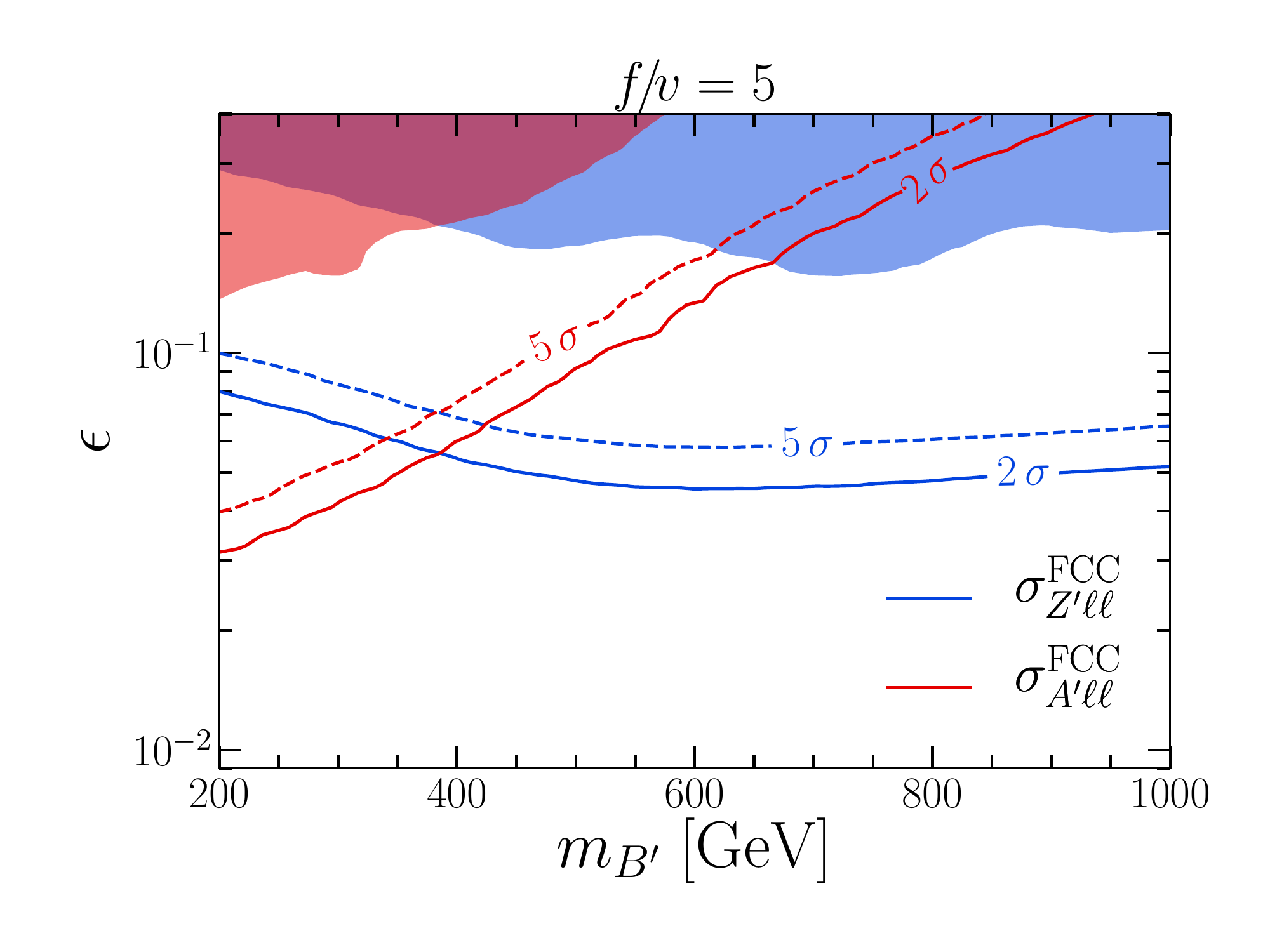}
\includegraphics[trim={9mm 8mm 9mm 9mm},clip,width=0.49\textwidth]{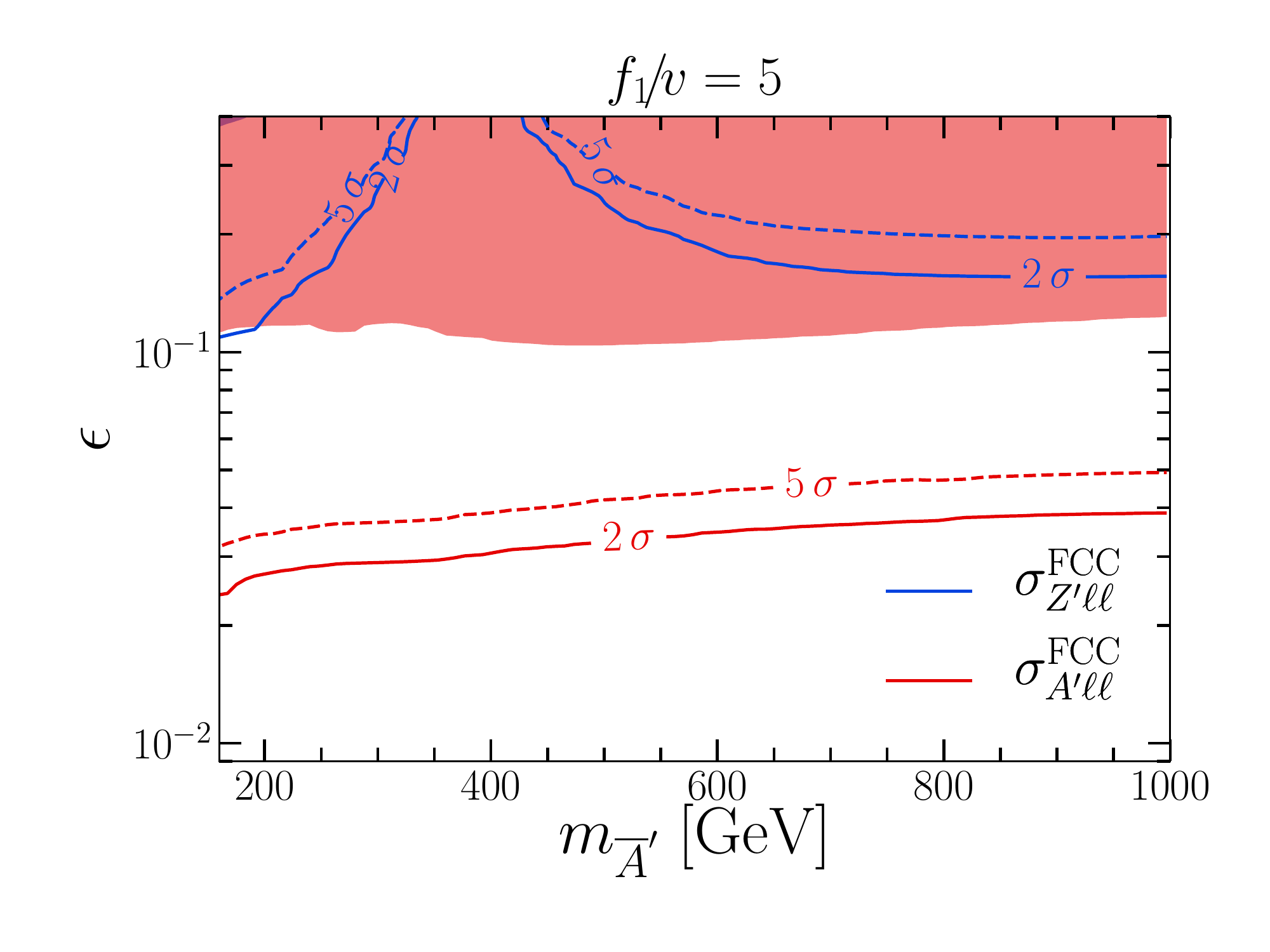}
\caption{Projected $\sigma$ contours of the $Z'$ (blue) and $A'$ (red) at a FCC hadron machine for a luminosity of 3000~fb$^{-1}$ for $f_{(1)}/v=3$(5) on the top (bottom) row. The THPM model is on the left and the T2HDM is on the right. The regions excluded by current LHC searches for the twin bosons are shaded in the corresponding color.}
\label{fig:futuresensitivityFCC}
\end{figure}

Now that the constraints on this framework have been mapped out, we 
determine the discovery and exclusion reach of the HL-LHC and a 100~TeV 
hadron collider. We estimate the sensitivity of future dilepton searches 
by a simple scaling up of existing results. As an example, if a run-II 
dilepton resonance search at a given mass can exclude a signal cross 
section of $\sigma_{S,13}$ at 95\% confidence level, then a similar 
search at a 100~TeV collider will be sensitive to a cross section 
$\sigma_{S,100}$ at the same confidence level, given by
 \beq
\sigma_{S,100}=\sigma_{S,13}\sqrt{\frac {\sigma_{\text{BG}100~{\rm TeV}}}{\sigma_{\text{BG}13~{\rm TeV}}}}\sqrt{\frac {\mathcal{L}_{13}}{\mathcal{L}_{100}}},
 \eeq
 where the square roots contain the ratios of the background cross 
sections at the two colliders, as well as the ratio of the luminosities 
of the two searches. We calculate the background cross sections at 13 
and 100~TeV at parton level in {\sc MadGraph}. Our results for 
$f_{(1)}/v=3$ and 5 are shown in Fig.~\ref{fig:futuresensitivityHLLHC} 
for the HL-LHC and Fig.~\ref{fig:futuresensitivityFCC} for the FCC-hh 
with a luminosity of $3000~{\rm fb}^{-1}$, the THPM model on the left 
and the T2HDM on the right. As in the collider constraints of the 
previous section, we see that in the THPM model the sensitivity to $A'$ 
falls off as $m_{B'}$ increases, while the sensitivity to $Z'$ persists. 
Conversely, in the T2HDM the $A'$ sensitivity falls off only slowly with 
$m_{\overline{A}'}$, while the sensitivity to the $Z'$ is much weaker 
across the entire parameter range.

The result is that in the THPM model, it is typically the $Z'$ channel 
that can be used to improve the limits on $\epsilon$ at the LHC and FCC, 
gaining a factor of a few over the constraints shown in 
Fig.~\ref{fig:constraints}, whereas in the T2HDM the $A'$ drives the 
sensitivity. This must be considered quite impressive, since for these 
resonances both the production cross section and the branching ratio 
into SM states scale roughly as $\epsilon^{2}$.

In the models we consider, the vector boson sector can be specified by 
three parameters: $f_{(1)}/v$, $\epsilon$, and either $m_{B'}$ or 
$m_{\overline{A}'}$. Therefore, measuring four or more independent 
observables can confirm the underlying structure. The observables in 
question can be the masses of $A'$ and $Z'$ and their resonant dilepton 
production rates, $\sigma_{\ell\ell,A'}$ and $\sigma_{\ell\ell, Z'}$. 
This is a goal that could be pursued at the LHC alone. Both vectors 
would appear as resonances in the dilepton spectrum. Then, measuring 
both masses and any one event rate would completely specify the parameters of the model. 
A measurement of the other rate then provides a test of the theory.

From Fig.~\ref{fig:futuresensitivityHLLHC}, we see that this is possible 
at the LHC for the THPM model. Although some part of the parameter space 
where both the $A'$ and the $Z'$ can be discovered is already ruled out by 
the current bounds on the $Z'$, a sizable region remains. In the T2HDM, 
however, this is not the case. Although the LHC has excellent 
sensitivity to the $A'$, the regions where the $Z'$ can be discovered at 
the HL-LHC are already ruled out by current bounds on the $A'$.

Before a FCC-hh machine begins taking data, it is very likely the tunnel 
will be used for a lepton collider. The FCC-ee is projected to measure 
some Higgs couplings to better than 1\%, an order of magnitude beyond 
the HL-LHC~\cite{Dawson:2013bba}. In particular, the coupling between 
the Higgs and $Z$ bosons may be measured to 0.3\% at 95\% 
confidence~\cite{Gomez-Ceballos:2013zzn}. We therefore expect that 
before the FCC-hh begins taking data, precise measurements of the Higgs 
couplings will already have determined $v/f_{(1)}$ to high accuracy. In both the THPM model and the T2HDM, the hadron machine can then completely specify 
the twin vector boson sector by measuring the mass and dilepton event 
rate of just one of the vectors, either the $A'$ or the $Z'$. In each of these models, the mass 
and dilepton event rate of the second vector are then predicted, and a 
targeted search may be made to test these theories. From 
Fig.~\ref{fig:futuresensitivityHLLHC} we see that in the THPM model, 
there are regions of parameter space where both the $A'$ and $Z'$ can be 
discovered at the FCC-hh. However, in the T2HDM the regions where the 
$Z'$ can be discovered are already ruled out by the current LHC bounds 
on the $A'$, so the model cannot be tested with this approach.


\section{Conclusions\label{s.Con}}

MTH models offer a simple solution to the little hierarchy problem 
without introducing new states charged under the SM gauge groups. The 
MTH framework predicts both a twin photon and a twin $Z$ boson. These 
states can interact with the SM through kinetic mixing between the 
hypercharge gauge boson of the SM and its mirror counterpart. If the 
twin photon is massive, this mixing can be sizable without violating the 
current experimental bounds. This portal can then be exploited by the 
LHC and future colliders to discover the twin vector bosons.

We have determined the bounds on the $A'$ and $Z'$ vector boson masses 
in a model in which the twin hypercharge gauge boson has a Proca mass, 
and also in a twin version of the two-Higgs-doublet model. In most of 
the parameter space, LHC searches for neutral gauge bosons constrain the 
mixing parameter $\epsilon \lesssim 0.1$ for both the THPM and T2HDM. 
The HL-LHC and a 100 TeV collider with 3000 $\text{fb}^{-1}$ of 
luminosity can improve on the current bounds by a factor of a few in 
most of the parameter space. In the THPM, in some regions of parameter space 
the HL-LHC or a future 100 TeV machine can discover both the neutral 
twin sector vector bosons. By measuring both the masses of these 
particles and the event rates into dilepton final states these colliders 
can test the THPM model. However, in the case of the T2HDM, the current 
LHC bounds on the $A'$ already exclude the parameter space in which the 
HL-LHC or FCC-hh would be expected to discover the $Z'$, and therefore we 
cannot test the model with this approach.

\section*{Acknowledgements}

We are grateful to Aqeel Ahmed and Anson Hook for helpful discussions. 
Z.C. is supported in part by the National Science Foundation under Grant 
Number PHY-1620074. Z.C. would like to thank the Fermilab Theory Group for 
hospitality during the completion of this work. Z.C.'s stay at Fermilab 
was supported by the Fermilab Intensity Frontier Fellowship and by the 
Visiting Scholars Award \#17-S-02 from the Universities Research 
Association. The research of C.K. is supported by the National Science 
Foundation Grant Number PHY-1620610. S.N. is supported by Vrije 
Universiteit Brussel through the Strategic Research Program ``High 
Energy Physics" and also supported by FWO under the EOS-be.h project n. 
30820817. C.B.V. is supported by Department of Energy Grant Number 
DE-SC-0009999.

\appendix
\section{Model Details and Diagonalization\label{a.Diag}}

The SM and twin sectors couple through the hypercharge portal. 
Consequently the SM photon and $Z$ boson mix with their twin 
counterparts, and the physical eigenstates are linear combinations of 
these states. In this appendix we diagonalize the Lagrangian for these 
vector bosons and determine the physical eigenstates. The full process 
of diagonalization involves several steps, which we take in turn. First, 
we define twin sector states analogous to the familiar photon and $Z$ 
boson in the SM,
 \begin{equation}
\overline{Z}_{\mu}^{\p} \equiv c_WW_\mu^{\p 3}-s_WB_\mu^\p, 
\ \ \ \  
\overline{A}_{\mu}^\p \equiv s_WW_\mu^{\p 3}+c_WB_\mu^\p,
 \end{equation}
 with $\cos\theta_W\equiv c_W$ and $\sin\theta_W\equiv s_W$, where 
$\theta_W$ is the weak mixing angle. For each of the models treated in 
the text, the Proca mass model (THPM) and twin two-Higgs-doublet model 
(T2HDM), this leads to a mass mixing matrix,
 \begin{align}
\frac12 \left(\overline{A}_{\mu}^\p\;\; \overline{Z}_{\mu}^{\p} \right)\left( \begin{array}{cc}
c_W^2m^2_{B'} & -s_W c_W m^2_{B'}\\
-s_W c_W m^2_{B'} &m^2_{\overline{Z}^\p}+ s_W^2m^2_{B'}
\end{array}\right)\left(\begin{array}{c}
\overline{A}_{\mu}^\p\\
\overline{Z}_{\mu}^{\p}
\end{array} \right),& \ \ \ \ \text{THPM\label{e.massmatrixPMTH}}\\
\frac12 \left(\overline{A}_{\mu}^\p\;\; \overline{Z}_{\mu}^{\p} \right)\left( \begin{array}{cc}
m_{\overline{A}'}^2 & m_{\overline{A}'}^2/t_{2W}\\
m_{\overline{A}'}^2/t_{2W} &m^2_{\overline{Z}^\p}+ m_{\overline{A}'}^2/t_{2W}^2
\end{array}\right)\left(\begin{array}{c}
\overline{A}_{\mu}^\p\\
\overline{Z}_{\mu}^{\p}
\end{array} \right).& \ \ \ \ \text{T2HDM}
\label{e.massmatrix2HDM}
 \end{align}
 We move to the diagonal basis, denoted with the subscript $0$, through the 
rotation matrix
 \begin{equation}
\left(\begin{array}{c}
Z_{0\mu}^\p\\
A_{0\mu}^{\p}
\end{array} \right)=\left(\begin{array}{cc}
\cos\phi & -\sin\phi\\
\sin\phi & \cos\phi
\end{array} \right)\left(\begin{array}{c}
\overline{A}_{\mu}^\p\\
\overline{Z}_{\mu}^{\p}
\end{array} \right).\label{e.massmatrix}
 \end{equation}
 The eigenvalues are given by
 \begin{align}
m^2_{Z'_0,A'_0}&=\frac{m^2_{\overline{Z}^\p}+m^2_{B'}}{2}\pm\frac12\sqrt{\left( m^2_{B'}-m^2_{\overline{Z}^\p}\right)^2+4s_W^2m^2_{B'}m^2_{\overline{Z}^\p}}\, , \ \ \ \ \text{THPM}\\
m^2_{Z'_0,A'_0}&=\frac{m^2_{\overline{Z}^\p}s_{2W}^2+m^2_{\overline{A}^\p}}{2s_{2W}^2}\pm \frac12\sqrt{\left(m^2_{\overline{Z}^\p}+c_{4W}\frac{m^2_{\overline{A}^\p}}{s_{2W}^2} \right)^2+4\frac{m^4_{\overline{A}^\p}}{t_{2W}^2}}, \ \ \ \ \text{T2HDM}
 \end{align}
 and the mixing angle is
 \begin{align}
 \sin2\phi=\frac{m^2_{B'}s_{2W}}{\sqrt{\left( m^2_{B'}-m^2_{\overline{Z}^\p}\right)^2+4s_W^2m^2_{B'}m^2_{\overline{Z}^\p}}}~,
 \end{align}
 for the THPM model. 
 The T2HDM has
 \begin{align}
 \sin2\phi&= -\frac{m_{\overline{A}'}^2s_{4W}}{\sqrt{\left(m^2_{\overline{Z}^\p}s_{2W}^2+c_{4W}m^2_{\overline{A}^\p} \right)^2+s_{4W}^2m^4_{\overline{A}^\p}}}~.
 \end{align}
 In the discussion that follows, the results are expressed in terms of 
the angle $\phi$. Therefore, all the results we obtain apply to both 
models, with the relation to the model parameters determined by the 
equations above.

The Lagrangian for the neutral vector bosons in the visible and twin 
sectors takes the form
 \begin{align}
\mathcal{L}\supset&-\frac{1}{4}{W_{0\mu\nu}}{W}_0^{\mu\nu}-\frac{1}{4}{A_{0\mu\nu}^{\p}}{A^{\p}}_0^{\mu\nu}-\frac{1}{4}{Z_{0\mu\nu}^{\p}}{Z^{\p}}_0^{\mu\nu}-\frac{1}{4}{B_{0\mu\nu}}{B}_0^{\mu\nu}+\frac{m_{Z_0}^2}{2}Z_{0\mu}Z^\mu_0\nonumber\\
&+\frac{\epsilon\, c_T}{2 } B_{0\mu\nu}Z^{\p 0\mu\nu}-\frac{\epsilon\, s_T}{2 }B_{0\mu\nu}A^{\p 0\mu\nu}+\frac{m^2_{A'_0}}{2}A^{\p}_{0\mu}A_0^{\p\mu}+\frac{m^2_{Z'_0}}{2}Z^{\p}_{0\mu}Z_0^{\p\mu},\label{e.twokineticlags}
 \end{align}
 where we have used the definitions  
 \begin{equation}
c_T\equiv {\cos(\theta_W-\phi)}, \qquad s_T\equiv {\sin(\theta_W-\phi)}.
 \end{equation}
 Here
 \beq
X_{\mu\nu}\equiv\partial_\mu X_\nu-\partial_\nu X_\mu,
 \eeq
 for each Abelian vector $X$, with the non-Abelian generalization used 
when appropriate. Note that for the THPM model, the limit $m_{B'}\gg 
m_{\overline{Z}'}$ leads to $\phi \approx \theta_W$. Then, in this limit 
$c_T\approx 1$ and $s_T\approx 0$. Therefore, we expect $A'$ to decouple 
from the SM for large $m_{B'}$. In the T2HDM, the limit 
$m_{\overline{A}'}=s_{2W}m_{\overline{Z}'}$ leads to 
$\phi=\theta_W-\pi/2$. In this limit $c_T = 0$ and $s_T = -1$, meaning 
that $Z'$ decouples from the SM for this particular ratio of the twin 
sector VEVs.

 Returning to the Lagrangian in Eq.~\eqref{e.twokineticlags}, we note 
that we may completely unmix all the kinetic terms by the following 
transformation
 \beq
\left(\begin{array}{c}
B_{0\mu}\\
Z^\p_{0\mu}\\
A^\p_{0\mu}
\end{array} \right)=\left(\begin{array}{ccc}
1 & \displaystyle c_T\epsilon(1+\alpha)& \displaystyle -\epsilon s_T(1+\alpha) \\
0 & \displaystyle 1+c_T^2\alpha & -c_T s_T\alpha\\
0 &-c_T s_T\alpha& \displaystyle 1+s_T^2\alpha
\end{array} \right)\left(\begin{array}{c}
{B}_{1\mu}\\
{Z}^\p_{1\mu}\\
{A}^\p_{1\mu}
\end{array} \right),
 \eeq
where
\beq
\alpha=-1+\frac{1}{\sqrt{1-\epsilon^2}}.
\eeq
 In the absence of kinetic mixing the eigenstates in the visible sector
are given by,
 \begin{align}
Z_{0\mu}&\equiv c_W W^3_\mu-s_W B_{0\mu},	
&A_{0\mu}&\equiv s_W W^3_\mu+c_W B_{0\mu}.	
 \end{align}
 We define 
 \begin{align}
Z_{1\mu}&\equiv c_W W^3_\mu-s_W {B}_{1\mu},
&A_{1\mu}&\equiv s_W W^3_\mu+c_W {B}_{1\mu}.
 \end{align}
 We can express $Z_{0\mu}$ and $A_{0\mu}$ in terms of $Z_{1\mu}$, $A_{1\mu}$,
$Z'_{1\mu}$ and $A'_{1\mu}$,
 \begin{align}
Z_{0\mu}&=Z_{1\mu}-\epsilon s_W(1+\alpha)\left(c_T{Z}^\p_{1\mu}-s_T{A}^\p_{1\mu} \right),	\\
A_{0\mu}&=A_{1\mu}+\epsilon c_W(1+\alpha)\left(c_T{Z}^\p_{1\mu}-s_T{A}^\p_{1\mu} \right).
 \end{align}
 Expressed in terms of $Z_{1\mu}$, $A_{1\mu}$, $Z'_{1\mu}$ and 
$A'_{1\mu}$, the Lagrangian is given by
 \begin{align}
\mathcal{L}\supset&-\frac{1}{4}Z_{1\mu\nu}Z_1^{\mu\nu}-\frac{1}{4}A_{1\mu\nu}A_1^{\mu\nu}-\frac{1}{4}{A}_{1\mu\nu}^{\p}{A}_1^{\p\mu\nu}-\frac{1}{4}{Z}_{1\mu\nu}^{\p}{Z}_1^{\p\mu\nu}\nonumber\\
&+\frac{m^2_{Z'_0}}{2}\left[(1+c_T^2\alpha)Z_{1\mu}'-c_Ts_T\alpha A_{1\mu}'\right]^2\nonumber\\
&+\frac{m^2_{A'_0}}{2}\left[ (1+s_T^2\alpha)A_{1\mu}'-c_Ts_T\alpha Z_{1\mu}' \right]^2\nonumber\\
&+\frac{m_{Z_0}^2}{2}\left[Z_{1\mu}-s_W\epsilon(1+\alpha)(c_TZ_{1\mu}'-s_TA_{1\mu}')\right]^2.
 \end{align}
 This leads to a mass matrix with characteristic equation
 \beq
(\lambda-m_{Z_0}^2)(\lambda-m^2_{Z'_0})(\lambda-m^2_{A'_0})+\epsilon^2\lambda(\lambda-c_W^2m^2_{Z_0})(\lambda-s_T^2m^2_{Z'_0}-c_T^2m^2_{A'_0})=0.
 \eeq
 Finding the exact mass eigenvalues from this equation is not simple, 
but we note a few features. First, in the limit $\epsilon\to 0$ the 
eigenvalues are $m^2_{A'_0}$, $m^2_{Z'_0}$, and $m_{Z_0}^2$, as 
expected. Direct inspection of the eigenvalue equation also shows that 
the leading correction to these values arises at order $\epsilon^2$. 
These corrections are given by,
 \begin{align}
m_Z^2&=m_{Z_0}^2+s_W^2\epsilon^2m_{Z_0}^2\left(\frac{c_T^2m_{Z_0}^2}{m_{Z_0}^2-m^2_{Z'_0}}+ \frac{s_T^2m_{Z_0}^2}{m_{Z_0}^2-m^2_{A'_0}}\right), \label{e.massesZ}\\
m_{Z'}^2&=m^2_{Z'_0} +\epsilon^2c_T^2m^2_{Z'_0}\frac{m^2_{Z'_0}-m_{Z_0}^2c_W^2}{m^2_{Z'_0}-m_{Z_0}^2},\\
m_{A'}^2&=m^2_{A'_0}+\epsilon^2s_T^2m^2_{A'_0}\frac{m^2_{A'_0}-m_{Z_0}^2c_W^2}{m^2_{A'_0}-m_{Z_0}^2} .\label{e.masses}
 \end{align}
  These formulae break down near mass degeneracies. At these points the 
mass matrix can be diagonalized numerically. One can also work out the 
corresponding eigenvectors and hence the similarity transform to the 
diagonal basis at order $\epsilon^2$.

 However, we can see the leading effects of the mixing by just working 
to linear order in $\epsilon$. We obtain 
 \begin{align}
Z_{0\mu}=&Z_{\mu}\left[1-\frac{\epsilon^2 m_{Z_0}^4s_W^2}{2}\left(\frac{s_T^2}{(m_{A_0^\p}^2-m_Z^2)^2}+\frac{c_T^2}{(m_{Z_0^\p}^2-m_{Z_0}^2)^2}\right)\right]\nonumber\\
&-\epsilon s_W\left[Z'_{\mu}\frac{c_T m^2_{Z'_0}}{m^2_{Z'_0}-m^2_{Z_0}}-A'_{\mu}\frac{s_T m^2_{A'_0}}{m^2_{A'_0}-m^2_{Z_0}} \right],\label{e.Ztrans}\\
A_{0\mu}=&A_\mu+\epsilon c_W\left(c_TZ'_{\mu}-s_TA'_{\mu} \right),\label{e.Atrans}\\
Z'_{0\mu}=& Z'_{\mu}+Z_{\mu}\frac{\epsilon s_W c_T m_{Z_0}^2}{m_{Z'_0}^2-m_{Z_0}^2}~, \\
A'_{0\mu}=& A'_{\mu}-Z_{\mu}\frac{\epsilon s_W s_T m_{Z_0}^2}{m_{A'_0}^2-m_{Z_0}^2} ~.
 \end{align}
 We have kept effects of order $\epsilon^2$ in the expression for the 
$Z$, because the correction to the $Z$ couplings to this order is 
important to obtain the constraints in Sec.~\ref{sec:constraints}. The 
results in the body of the paper are based on a numerical analysis, so 
the formulae in this appendix are intended only to provide a qualitative 
understanding.

\section{Vector to Fermion Couplings\label{a.VFcouplings}}

 In this section we record the couplings of the gauge bosons to the 
fermions of the two sectors, and determine their partial widths. The 
visible photon's couplings are simple,
 \begin{equation}
g_{A f\bar{f}}=gs_W Q=eQ, \ \ \ \ g_{A f'\bar{f}'}=0~.
 \end{equation}
 The couplings of the $Z$, $A^\p$ and $Z^\p$ bosons to fermions and twin 
fermions can be parametrized as,
 \begin{align}
{\cal L}_\text{Int}&= g_{Zf\bar{f}} \bar{f}\gamma^\mu{f} Z_{\mu}+ g_{A^\p f\bar{f}} \bar{f}\gamma^\mu{f}A^\p+ g_{Z^\p f\bar{f}} \bar{f}\gamma^\mu{f}Z^\p 		\notag\\
&+g_{Zf^\p \bar{f^\p }}\bar{f^\p }\gamma^\mu f^\p  Z_{\mu} +g_{A^\p f^\p \bar{f^\p }}\bar{f^\p }\gamma^\mu f^\p  A^\p+g_{Z^\p f^\p \bar{f^\p }}\bar{f^\p }\gamma^\mu f^\p  Z^\p.
 \end{align}
 The couplings of the massive neutral gauge bosons to the SM fermions 
are given by,
 \begin{align}
g_{Z f\bar{f}}=&\frac{g}{c_W}\left\{\left(T^3- Q s_W^2\right)\left[1-\frac{\epsilon^2 m_{Z_0}^4s_W^2}{2}\left(\frac{s_T^2}{(m_{A_0^\p}^2-m_Z^2)^2}+\frac{c_T^2}{(m_{Z_0^\p}^2-m_{Z_0}^2)^2}\right)\right]\right.\nonumber\\
&\left.\phantom{AA}+Y \epsilon^2 m_{Z_0}^2s_W^2\left(\frac{s_T^2}{m_{A_0^\p}^2-m_Z^2}+\frac{c_T^2}{m_{Z_0^\p}^2-m_Z^2}\right)\right\},\\
g_{Z^\p f\bar{f}}=&-\frac{g}{c_W}\epsilon s_W c_T \left[\left(T^3- Q s_W^2\right)\frac{m_{Z_0}^2}{m_{Z^\p_0}^2-m_{Z_0}^2}-Y \right],\\
g_{A^\p f\bar{f}}=&\frac{g}{c_W}\epsilon s_W s_T\left[\left(T^3- Q s_W^2\right)\frac{m_{Z_0}^2}{m_{A^\p_0}^2-m_{Z_0}^2}-Y\right].
 \end{align}
 The couplings of the massive neutral gauge bosons to the fermions of the 
twin sector are
 \begin{align}
g_{Z f^\p\bar{f^\p}}=&\frac{-g}{c_W}\epsilon m_{Z_0}^2s_W\left\{\left[\cos{\phi}\left(T^3- Q s_W^2\right)+Q s_Wc_W\sin{\phi } \right]\frac{ s_T}{m_{A^\p_0}^2-m_{Z_0}^2}\right.\nonumber\\
&\left.\phantom{AAAAA}+\left[\sin{\phi}\left(T^3- Q s_W^2\right)-Q s_Wc_W\cos{\phi } \right]\frac{ c_T}{m_{Z^\p_0}^2-m_{Z_0}^2}\right\},\\
g_{Z^\p f^\p \bar{f^\p}}&=\frac{-g}{c_W}\left[\sin\phi\left(T^3- Q s_W^2\right)-Q s_Wc_W\cos\phi \right],\\
g_{A^\p f^\p\bar{f^\p}}&=\frac{g}{c_W}\left[\cos\phi\left(T^3- Qs_W^2\right)+Q s_Wc_W\sin\phi \right] .
 \end{align}


In calculating the branching fractions of the various gauge bosons the 
following formulae are of use. The decay width of the vector boson $V(Z, 
A^\p,Z^\p)$ to fermions is given by,
 \beq
\Gamma(V\to ff)= \frac{N_c  m_V}{24\pi }\sqrt{1-4\frac{m_{f}^2}{m_V^2}}\left[\left({g_L^f}^2+{g_R^f}^2\right)\left(1-\frac{m_{f}^2}{m_V^2}\right)+6g^f_L g^f_R\frac{m_{f}^2}{m_V^2}\right]
 \eeq
 The neutral vector bosons also have a decay width to $Zh$, where $h$ 
denotes the Higgs. This contribution to the decay width arises from the 
$hZ_0Z_0$ and $h\overline{Z}'\overline{Z}'$ terms in the Lagrangian. 
This leads to the terms $g_{A'Zh}A'_\mu Z^\mu h$ and $g_{Z'Zh}Z'_\mu 
Z^\mu h$ with
 \begin{align}
g_{A'Zh}=&2\epsilon s_W\frac{m_{Z_0}^2}{v_\text{EW}}\left(1-\frac{v_\text{EW}^2}{2f_{(1)}^2} \right)\left[s_T\frac{m_{A^\p_0}^2+ m_{Z_0}^2\cos^2\phi}{m_{A^\p_0}^2-m_{Z_0}^2}+c_T\frac{m_{Z_0}^2\cos\phi\sin\phi}{m_{Z^\p_0}^2-m_{Z_0}^2} \right],\\
g_{Z'Zh}=&-2\epsilon s_W\frac{m_{Z_0}^2}{v_\text{EW}}\left(1-\frac{v_\text{EW}^2}{2f_{(1)}^2} \right)\left[c_T\frac{m_{Z^\p_0}^2+ m_{Z_0}^2\sin^2\phi}{m_{Z^\p_0}^2-m_{Z_0}^2}+s_T\frac{m_{Z_0}^2\cos\phi\sin\phi}{m_{A^\p_0}^2-m_{Z_0}^2} \right].
\end{align}
 The decay widths are then given by
 \begin{align}
&\Gamma(V\to Z h)= \frac{g_{V h Z}^2}{192\pi}\frac{m_{V}}{m_Z^2}\lambda\left(1,x_Z,x_h\right)\left[\lambda^2(1,x_Z,x_h)+12x_Z\right],\notag\\&
\lambda(a,b,c)=\sqrt{a^2+b^2+c^2-2ab-2bc-2ac},~~x_Z=\frac{m^2_Z}{m^2_{V}},~~x_h=\frac{m_h^2}{m_{V}^2}.
 \end{align}
 The decay of a vector $V$ into a pair of visible $W$ bosons arises from 
the usual $Z_0WW$ and $A_0 WW$ vertices in the visible sector and using 
Eqs.~\eqref{e.Ztrans} and \eqref{e.Atrans}. The width is then given by,
 \begin{align}
&\Gamma(V\to WW)= f(k) g_{V^\p WW}^2\frac{m_{V}}{192\pi},\notag\\&
f(k)=\left(1-4k\right)^{\frac{3}{2}}\left(1+20k+12k^2\right)k^{-2},~~k=\frac{m_W^2}{m_{V}^2}.
 \end{align}
 Decays of the $Z'$ into $W'W'$ have the same form with the coupling arising from the $\overline{Z}'W'W'$ and $\overline{A}' W'W'$ vertices.





\bibliographystyle{JHEP}
\bibliography{bib_twin}{}
\end{document}